\documentclass[aps, prc, preprint, groupedaddress]{revtex4-2}
\usepackage[utf8]{inputenc}
\usepackage{setspace,amsmath, amssymb}
\usepackage{graphicx}
\usepackage{titlesec}
\usepackage{subcaption}
\usepackage{geometry}

\usepackage[colorlinks=true, linkcolor=blue, citecolor=blue, urlcolor=blue]{hyperref}
\hypersetup{
	pdfdisplaydoctitle=true,
	pdfstartview={FitH},
}
\renewcommand{\thesection}{\arabic{section}}
\renewcommand{\thesubsection}{\arabic{section}.\arabic{subsection}}

\titleformat{\section}
{\normalfont\large\bfseries}
{\thesection}{1em}{}

\titleformat{\subsection}
{\normalfont\bfseries}
{\thesubsection}{1em}{}

\titleformat{\subsubsection}
{\normalfont\itshape}
{\thesubsubsection}{1em}{}

\begin{document}
	\title{Comparing theoretical descriptions of Drell–Yan angular coefficients}
	
	\author{Nikita Gramotkov}
	\affiliation{Bogoliubov Laboratory of Theoretical Physics, JINR, Dubna, Russia}
	\affiliation{Department of Physics, Moscow State University, Moscow, Russia}
	
	\author{Oleg Teryaev}
	\affiliation{Bogoliubov Laboratory of Theoretical Physics, JINR, Dubna, Russia}
	\affiliation{Department of Physics, Moscow State University, Moscow, Russia}
	
	\begin{abstract}
	 A quantitative comparison of the predictions of three theoretical descriptions (TMD, pQCD and geometrical) of the angular coefficients in Drell-Yan $Z$-boson production with new experimental data on angular coefficients obtained by the ATLAS, CMS and LHCb collaborations has been performed. In the central rapidity region at low-$Q_T$, the TMD calculation provides a better description of the $A_0$ and $A_2$ coefficients among the descriptions considered. The differences between the approaches are less pronounced for $A_1, A_3$ and $A_4$ coefficients. The results identify $A_0, A_2$ and the Lam-Tung relation as the most sensitive observables at low-$Q_T$ and illustrate the domains of applicability of the three descriptions. 
	\end{abstract}
	
\maketitle	
\section{Introduction}
	The Drell--Yan (DY) process~\cite{Drell:1970yt,Matveev:1969zz,Christenson:1970um}, 
	where a quark-antiquark pair  annihilates into a dilepton in hadronic collisions, remains one of the key tools for studying quantum chromodynamics (QCD). Measuring the angular distributions of final-state leptons provides detailed information about hadron structure and parton dynamics. These distributions can be decomposed into angular functions, which are related to eight elements of gauge boson spin-density matrix~\cite{Gavrilova:2019jea}. Coefficients of these functions are sensitive to various aspects of parton distributions, including transverse-momentum effects and spin correlations.
	The standard expression of angular distributions for the DY process in the rest frame of a lepton pair (the Collins-Soper (CS) frame), widely used in theoretical and experimental work, has the following form:
	\begin{align}
		\nonumber
		\frac{dN}{d\Omega} &= \frac{3}{16\pi} \Bigg[ 1 + \cos^2\theta + \frac{A_0}{2} (1 - 3\cos^2\theta) + A_1 \sin 2\theta \cos\phi + \frac{A_2}{2} \sin^2\theta \cos 2\phi \\ 
		&+ A_3 \sin\theta \cos\phi + A_4 \cos\theta + A_5 \sin^2\theta \sin 2\phi + A_6 \sin 2\theta \sin\phi + A_7 \sin\theta \sin\phi \Bigg],
		\label{eq:angular_distributions}
	\end{align}
	here $\theta$ and $\phi$ are polar and azimuthal angles of outgoing leptons, and the coefficients $A_0 - A_7$ are functions of kinematic variables and carry information about the structure of hadrons and the dynamics of partons.
	\begin{align}
		 \nonumber
		\frac{dN}{d\Omega}
		&=
		\frac{d\sigma}{d\Omega d^4q}
		\left(\frac{d\sigma}{d^4q}\right)^{-1}
		\\ \nonumber
		&= \frac{3}{8\pi(2W_T + W_L)}
		\Big[
		g_T W_T + g_L W_L + g_\Delta W_\Delta + g_{\Delta\Delta} W_{\Delta\Delta} \\
		&+ g_{TP} W_{TP} + g_{\nabla P} W_{\nabla P} + g_\nabla W_\nabla + g_{\Delta\Delta P} W_{\Delta\Delta P} + g_{\Delta P} W_{\Delta P}
		\Big],
		\label{eq:angular_distributions_Q2}
	\end{align}
	Here $q^{\mu}$ denotes the four-momentum of the dilepton pair, with $Q^2 = q^2$. The structure functions $W_i$, and consequently the angular coefficients $A_i \ (\text{or} \ g_i)$, depend on the dilepton kinematics; in the following, we focus on their $Q_T$ dependence for different values of $Q^2$ and rapidity $y$.
	A key theoretical benchmark for these angular distributions is the Lam-Tung relation~\cite{Lam:1978pu,Lam:1980uc}, which in the notation of eq.~(\ref{eq:angular_distributions}) predicts a specific linear correlation between the coefficients $A_0$ and $A_2$. This relation, analogous to the Callan-Gross relation in deep-inelastic scattering~(DIS), was expected to hold in leading-order QCD. However, significant violation was found in unpolarized DY experiments with pion beams~\cite{NA10:1986fgk,E615:1989bda} and LHC experiments on $Z$-boson production~\cite{ATLAS:2016rnf,CMS:2015cyj}. This violation can be attributed, in particular, to the Boer-Mulders function~\cite{Boer:1997nt,Boer:1999mm}, which describes the correlation between the quark transverse spin and its transverse momentum inside a hadron. This function directly generates the $\cos{2\phi}$ asymmetry and thus violates the Lam–Tung relation. 
	However, at small transverse momentum, fixed-order collinear calculations contain logarithmically singular contributions of the form $\ln( Q^{2}/Q_{T}^{2})$ which require an all-order resummation when $Q_T \ll Q$. Such a resummation was originally developed within the Collins-Soper-Sterman formalism~\cite{Collins:1981uk,Collins:1984kg} and is incorporated in the modern TMD factorization framework, where the transverse momentum of the partons are retained  explicitly~\cite{Tangerman:1994eh,Arnold:2008kf}. An alternative approach is the geometrical model~\cite{Peng:2015spa}, parametrising the transverse momentum in terms of the geometry of the partonic configuration. The three approaches represent different levels of theoretical description: fixed-order perturbative QCD, transverse-momentum-dependent factorization including power corrections, and an effective phenomenological geometrical model. While in collinear pQCD the Lam-Tung relation is violated only at NNLO, in TMD factorization it receives contribution from Boer-Mulders function and also from kinematic power corrections (KPCs) already at leading power (LP), and in geometrical model it follows directly from the acoplanarity between the quark and hadron planes. These three approaches describe DY angular distributions from different theoretical perspectives, but a direct comparison using the same experimental dataset has been performed only to a few of them. A direct comparison can show how the agreement with data changes between different kinematic regions and which mechanisms are relevant for the observed angular asymmetries. In this work, we compare the three descriptions for the coefficients $A_0 - A_4$  with the same experimental dataset on Z-boson production obtained by the ATLAS~\cite{ATLAS:2016rnf}, CMS~\cite{CMS:2015cyj} including recent results~\cite{CMS:2026amb}, and LHCb~\cite{LHCb:2022tbc} collaborations over a wide range of transverse momenta $Q_T$ and dilepton rapidities $y$. Particular attention is paid to the coefficients $A_0$, $A_2$ and to the Lam-Tung relation , which are sensitive to transverse momentum dynamics. The coefficients $A_1,A_3$ and $A_4$ are also considered, providing a comparison for angular structures with a different sensitivity to transverse momentum effects. We also examine how the agreement of the three approaches with the data changes across different kinematic regions.
\section{Theoretical approaches}
\subsection{Collinear factorization with small-$Q_T$ expansion}
The first theoretical approach considered in this paper is collinear factorization in the framework of perturbative QCD. The key element of this approach is the factorization of the hadronic tensor into parton structure functions and parton distribution functions (PDF), which takes the following form 
\begin{align}
	W(x_1, x_2) = \frac{1}{x_1 x_2} \sum_{a,b} \int_{x_1}^{1} dz_1 \int_{x_2}^{1} dz_2 \, \tilde{w}^{ab}(z_1, z_2) \, f_{a/H_1}\!\left(\frac{x_1}{z_1}\right) f_{b/H_2}\!\left(\frac{x_2}{z_2}\right),
	\label{eq:collinear_factorization}
\end{align}
where $W$ denotes hadronic structure functions, $\tilde{w}^{ab}$ - parton structure functions, $f_{a/H}$ - PDFs. This factorization formula can be calculated perturbatively order by order in the strong coupling $\alpha_s$.
In the collinear factorization approach, the angular coefficients are determined by the hadronic structure functions (\ref{eq:collinear_factorization}), which are related to the polarization of virtual boson as follows:
\begin{gather}
	\label{eq:collinear_coeffs}
	A_0 = \frac{2 W_L}{2 W_T + W_L}, \quad
	A_1 = \frac{2 W_\Delta}{2 W_T + W_L}, \quad
	A_2 = \frac{4 W_{\Delta\Delta}}{2 W_T + W_L}, \\ \nonumber
	A_3 = \frac{2 W_{\nabla_P}}{2 W_T + W_L}, \quad
	A_4 = \frac{2 W_{T_P}}{2 W_T + W_L}, \quad
	A_5 = \frac{2 W_{\Delta\Delta_P}}{2 W_T + W_L}, \\ \nonumber
	A_6 = \frac{2 W_{\Delta_P}}{2 W_T + W_L}, 
	A_7 = \frac{2 W_{\nabla}}{2 W_T + W_L}.
\end{gather}
This formalism was introduced in classical works~\cite{Lam:1980uc,Collins:1977iv} and was later applied to the low-$Q_T$ angular distribution in Ref.~\cite{Boer:2006eq}. However, when one tries to compute the full cross section in the low-$Q_T$ limit directly from eq.~(\ref{eq:collinear_factorization}), the individual hadronic structure functions $W_T, W_L$ etc., develop power divergences $~ 1/\rho^2$ (with $\rho^2 = Q^2_{T}/Q^2$) from initial-state gluon emission~\cite{Berger:2007si,Berger:2007jw}. This leads to an unphysical growth of the fixed-order cross section at $Q_T \rightarrow 0$. But its divergence cancels exactly in their ratios, which define the angular coefficients~(\ref{eq:collinear_coeffs}). Consequently, coefficients $A_i$ remain finite and well-defined in the collinear factorization approach, which makes them particularly suitable for comparison with experimental data at low-$Q_T$. In contrast, the $T$-odd structure functions, generated by absorptive parts of partonic amplitudes at $\mathcal{O}(\alpha^2_s)$, do not contain the leading $1/\rho^2$ singularity, although logarithmic terms are still present in their low-$Q_T$ expansion~\cite{Lyubovitskij:2024civ}. 
In recent papers~\cite{Lyubovitskij:2024civ,Lyubovitskij:2024jlb}, this formalism was extended to the low-$Q_T$ region. In particular, a systematic expansion of the structure functions in powers of $\rho^2$ up to NNLP was performed, including the electroweak contributions from $Z^0$-boson exchange~[see sec.IV in Ref.~\cite{Lyubovitskij:2024jlb}], that allows a direct comparison of theoretical predictions with precise experimental LHC data across a wide $Q_T$ range. LP results for hadronic structure functions in eq.~(\ref{eq:collinear_coeffs}) at this extension have the following form for $q\bar{q}$-subprocess:
\begin{align}
	\label{eq:W_qq}
	W^{\text{LP};q\bar{q}}_{\text{T}} &= \dfrac{g_{q\bar{q};1}}{C_{F}\rho^2 x^{0}_{1} x^{0}_{2}} \left[-C_{F}(2L_{\rho} + 3)q_1 (x^{0}_{1})\bar{q}_2 (x^{0}_{2}) + q_1 (x^{0}_{1})
	(P_{qq} \otimes \bar{q}_{2})(x^{0}_{2}) + (P_{qq} \otimes q_{1})(x^{0}_{1})\bar{q}_2 (x^{0}_2)\right] \\
	W^{\text{LP};q\bar{q}}_{\Delta} &= \dfrac{g_{q\bar{q};1}}{C_{F}\rho x^{0}_{1} x^{0}_{2}} \left[q_1 (x^{0}_{1})
	(\tilde{P}_{qq} \otimes \bar{q}_{2})(x^{0}_{2}) - (\tilde{P}_{qq} \otimes q_{1})(x^{0}_{1})\bar{q}_2 (x^{0}_2)\right]
\end{align}
with the identities:
\begin{gather}
	\nonumber
	W^{\text{LP};q\bar{q}}_{\text{T}} = \dfrac{1}{\rho^2} 	W^{\text{LP};q\bar{q}}_{\text{L}} = \dfrac{2}{\rho^2} 	W^{\text{LP};q\bar{q}}_{\Delta \Delta} = \dfrac{g_{q\bar{q};1}}{g_{q\bar{q};2}} W^{\text{LP};q\bar{q}}_{\text{T}_P}, \\
	W^{\text{LP};q\bar{q}}_{\Delta} = \dfrac{g_{q\bar{q};1}}{g_{q\bar{q};2}} W^{\text{LP};q\bar{q}}_{\nabla_P};
	\label{eq:W_qq_idnts}
\end{gather}
and for $q g$-subprocess:
\begin{align}
	W^{\text{LP};qg}_{\text{T}} &= \dfrac{2 g_{qg;1}}{\rho^2 x^0_1 x^0_2}q_1(x^0_1) \left(P^{+}_{qg} \otimes g_2\right) (x^0_2) \\ 
	W^{\text{LP};qg}_{\text{L}} &= \dfrac{2 g_{qg;1}}{x^0_1 x^0_2}q_1(x^0_1) \left(P^{-}_{qg} \otimes g_2\right) (x^0_2) \\
	W^{\text{LP};qg}_{\Delta} &= \dfrac{2 g_{qg;1}}{\rho x^0_1 x^0_2}q_1(x^0_1) \left(\tilde{P}_{qg} \otimes g_2\right) (x^0_2)
	\label{eq:W_qg}
\end{align}
with the following identities:
\begin{align}
	W^{\text{LP};qg}_{\text{T}} = \dfrac{g_{qg;1}}{g_{qg;2}} W^{\text{LP};qg}_{\text{T}_P}, \ \ W^{\text{LP};qg}_{\text{L}} = 2 W^{\text{LP};qg}_{\Delta\Delta}, \ \ W^{\text{LP};qg}_{\Delta} = \dfrac{g_{qg;1}}{g_{qg;2}} W^{\text{LP};qg}_{\nabla_P}.
	\label{eq:W_qg_idnts}
\end{align}
Where $(\mathcal{P}\otimes f)$ denotes a generalized convolution as following:
\begin{align}
	(\mathcal{P} \otimes f)(x) = \int\limits^{1}_{x} \dfrac{dz}{z} \mathcal{P}(z) f(\dfrac{x}{z}).
\end{align}
After substituting (\ref{eq:W_qq})-(\ref{eq:W_qg_idnts}) into the angular coefficients (\ref{eq:collinear_coeffs}), one can calculate them directly (see Appendix D in~\cite{Lyubovitskij:2024jlb} and Supplementary Material in~\cite{Lyubovitskij:2024civ} for detailed materials of NLP and NNLP analytical results).
\subsection{TMD factorization approach}
The second theoretical approach considered in this paper is based on a factorization that explicitly retains the partonic transverse momentum dependence, which is integrated over in collinear factorization. This provides a systematic resummation of large logarithms $\ln (Q^2/Q^2_{T})$ that occur in the region of small $Q_T$, where the perturbative calculations diverge. 
The leading power (LP) expansion of TMD factorization was developed in~\cite{Tangerman:1994eh,Arnold:2008kf}, where the hadron tensor has the form:
\begin{align}
	W^{\mu\nu}_{\mathrm{LP}} \sim \displaystyle\int d^2k_{1T}d^2k_{2T}\; \delta^{(2)}(\mathbf{q}_T - \mathbf{k}_{1T} - \mathbf{k}_{2T}) \; C_0 \; f_1(x_1,k_{1T}) \, f_1(x_2,k_{2T}),
\end{align}
here $f_1$ is unpolarized TMDPDF, $k_{1T},k_{2T}$ are the transverse momentum of partons. However, the LP expansion alone does not preserve gauge and frame invariance. Its frame dependence originates from neglected power-suppressed terms of order $(k_{T}/Q)^n$, which are required to restore these symmetry properties.
A recent formulation of TMD factorization~\cite{Vladimirov:2023aot} provides a systematic resummation of KPCs associated with the LP term. These corrections form a series of $k_T/Q$-suppressed terms associated with the LP contribution (cf. Ref.~\cite{Teryaev:2004df}). In coordinate space they appear as transverse derivatives of twist-two TMD distributions, corresponding to powers of the partonic transverse momenta $k_T$ in momentum space. Unlike the genuine higher-twist corrections, this series involves only twist-two TMD distributions and therefore introduces no additional nonperturbative functions. Their resummation restores gauge and Lorentz invariance while preserving the perturbative and nonperturbative content of the LP TMD factorization theorem. The resulting expression for the hadron tensor with KPCs takes the form: 
\begin{align}
	\nonumber
	W_{GG'}^{\mu\nu} &= \frac{1}{4N_c} C_0\left(\frac{Q^2}{\mu^2}\right) \int \frac{d^4y}{(2\pi)^4} e^{-i(yq)} \sum_{a,b}\sum_f \Bigg[
	\mathrm{Tr}\left(\gamma_G^\mu \bar{\Gamma}_b \gamma_{G'}^\nu \bar{\Gamma}_a\right)
	\widetilde{\Psi}_{f/p_1}^{[\Gamma_a]}(y;\mu,Q^2)
	\widetilde{\Psi}_{\bar{f}/p_2}^{[\Gamma_b]}(y;\mu,Q^2) \\ 
	&+ \mathrm{Tr}\left(\gamma_G^\mu \bar{\Gamma}_a \gamma_{G'}^\nu \bar{\Gamma}_b\right)
	\widetilde{\Psi}_{\bar{f}/p_1}^{[\Gamma_a]}(y;\mu,Q^2)
	\widetilde{\Psi}_{f/p_2}^{[\Gamma_b]}(y;\mu,Q^2) \Bigg] + \ldots,
\end{align}
where the correlators $\tilde{\Psi}$ are expressed in terms of ordinary twist-two TMD distributions: the unpolarized TMDPDF $f_1$ and the Boer-Mulders function $h^{\perp}_{1}$. Following Ref.~\cite{Piloneta:2024aac}, the complete set of angular coefficients within this factorization has the form
\begin{align}
	\Sigma_U = \dfrac{d\sigma}{d^4 q}, \quad A_{n} = \dfrac{\Sigma_n}{\Sigma_U}.
\end{align}
Expressions for the $\Sigma_i$ (where $i = U,0,1,2,3,4$) structure functions are the following
\begin{align}
	\label{eq:TMD_coeffs}
	\Sigma_U = \frac{4\pi\alpha_{\mathrm{em}}^2}{3N_c s}
	\sum_{q,G,G'} Q^4 \Delta_G^*\Delta_{G'} \,
	z_{+\ell}^{GG'} z_{+\bar q}^{GG'} \,
	\mathcal{C}\big[1,\, f_1 f_1\big]
\end{align}
\begin{align}
	\nonumber
	\Sigma_0 &= \frac{4\pi\alpha_{\mathrm{em}}^2}{3N_c s}
	\sum_{q,G,G'} Q^4 \Delta_G^*\Delta_{G'} \Bigg\{
	z_{+\ell}^{GG'} z_{+\bar q}^{GG'} \,
	\mathcal{C}\Bigg[\frac{(\mathbf{k}_1-\mathbf{k}_2)^2}{Q^2} - \frac{(k_1^2-k_2^2)^2}{\tau^2 Q^2},\, f_1 f_1\Bigg]
	\\
	&+ z_{+\ell}^{GG'} r_{+\bar q}^{GG'} \,
	\mathcal{C}\Bigg[\frac{(\mathbf{k}_1-\mathbf{k}_2)^2(k_1^2+k_2^2)}{2M^2 Q^2} - \frac{(k_1^2-k_2^2)^2}{M^2 Q^2}\Bigg(\frac12 - \frac{(\mathbf{k}_1\mathbf{k}_2)}{\tau^2}\Bigg),\, h_1^\perp h_1^\perp\Bigg]
	\Bigg\}
\end{align}
\begin{align}
	\Sigma_1 &\sim \mathcal{C}\left[\frac{|\mathbf{k}_1| - |\mathbf{k}_2|}{Q},\; f_1 f_1\right] 
	+ \mathcal{C}\left[\frac{|\mathbf{k}_1| - |\mathbf{k}_2|}{Q}\frac{(\mathbf{k}_1\cdot\mathbf{k}_2)}{M^2},\; h_1^\perp h_1^\perp\right], \\
	\Sigma_2 &\sim \mathcal{C}\left[\frac{(\mathbf{k}_1 - \mathbf{k}_2)^2}{Q^2},\; f_1 f_1\right] 
	+ \mathcal{C}\left[\frac{(\mathbf{k}_1 - \mathbf{k}_2)^2}{M^2},\; h_1^\perp h_1^\perp\right], \label{eq:TMD_S2} \\
	\Sigma_3 &\sim \mathcal{C}\left[\frac{|\mathbf{k}_1| - |\mathbf{k}_2|}{Q},\; \{f_1 f_1\}_A\right], \\
	\Sigma_4 &\sim \mathcal{C}\left[1,\; \{f_1 f_1\}_A\right],
	\label{eq:TMD_coeffs_end}
\end{align}
where $\tau^2 = Q^2 + Q^{2}_{T}$ and $\mathcal{C}$ denotes the convolution integral 
\begin{align}
	\nonumber
	\mathcal{C}[A,f_1 f_2] = &4p_1^+ p_2^- C_0\left(\frac{Q^2}{\mu^2}\right) 
	\int d\xi_1 d\xi_2 \int d^4 k_1 d^4 k_2 
	\delta^{(4)}(q - k_1 - k_2) \delta(k_1^2) \delta(k_2^2) \\
	&\delta(k_1^+ - \xi_1 p_1^+) \delta(k_2^- - \xi_2 p_2^-) 
	A \left( f_{q1}(\xi_1, \boldsymbol{k}_{1T}^2) f_{\bar{q}2}(\xi_2, \boldsymbol{k}_{2T}^2) 
	+ f_{\bar{q}1}(\xi_1, \boldsymbol{k}_{1T}^2) f_{q2}(\xi_2, \boldsymbol{k}_{2T}^2) \right).
\end{align}
The unpolarized TMDPDF $\tilde{f}_1$ is taken from the ART23~\cite{Moos:2023yfa} global fit, which includes data from fixed-target experiments, Tevatron and LHC
\begin{align}
	\tilde{f}_{1q}(x,b) = \int_x^1 \frac{dy}{y} \sum_{q'} C_{q \leftarrow q'}(y,b,\mu_{\mathrm{OPE}}) q_{q'}\left(\frac{x}{y}, \mu_{\mathrm{OPE}}\right) f_{\mathrm{NP}}^q(x,b),
\end{align}
where non-perturbative part has the following form
\begin{align}
	f_{\mathrm{NP}}^q(x,b) = \frac{1}{\cosh\big((\lambda_1^q(1-x) + \lambda_2^q x)b\big)}.
\end{align}
In a recent paper~\cite{Piloneta:2024aac} was suggested the following parametrisation for Boer-Mulders function
\begin{align}
	\tilde{h}_1^\perp(x,b) = \frac{2^{\alpha+1}}{\Gamma(\alpha+1)} \frac{N}{\cosh(\lambda b)} x \ln^\alpha(1/x),
	\label{eq:boer_mulders}
\end{align}
where the parameters extracted from the ATLAS data on the $A_2$ angular coefficient, yielding $\lambda = 0.2$ GeV (fixed), $N = -0.27^{+0.34}_{-0.12}$ and $\alpha = 9.4^{+5.4}_{-0.9}$.
We use the ART23 global extraction of the unpolarized TMDPDF~\cite{Moos:2023yfa}, which was also employed in Ref.~\cite{Piloneta:2024aac}. Using the same nonperturbative input as in Ref.~\cite{Piloneta:2024aac} makes the comparison less sensitive to differences in TMD parametrization. Within the TMD factorization with KPCs considered here, the coefficients $A_0 - A_4$ are described in terms of the unpolarized TMDPDF and the Boer-Mulders function. The discussion is restricted to the T-even coefficients, since the T-odd coefficients $A_5-A_7$ vanish under adopted assumptions and at the accuracy considered. 
\subsection{Geometrical approach}
The geometric approach developed in Refs.~\cite{Chang:2017kuv,Chang:2018pvk,Peng:2015spa,Teryaev:2005,Peng:2018tty,Peng:2019boj} provides a description of the DY angular distributions in terms of the geometry of the partonic process. Rather than calculating the angular coefficients directly from a factorization formula, this approach parametrizes them in terms of geometric functions characterizing the partonic configuration.  
\begin{figure}[h!]
	\centering
	\includegraphics[width=0.5\linewidth]{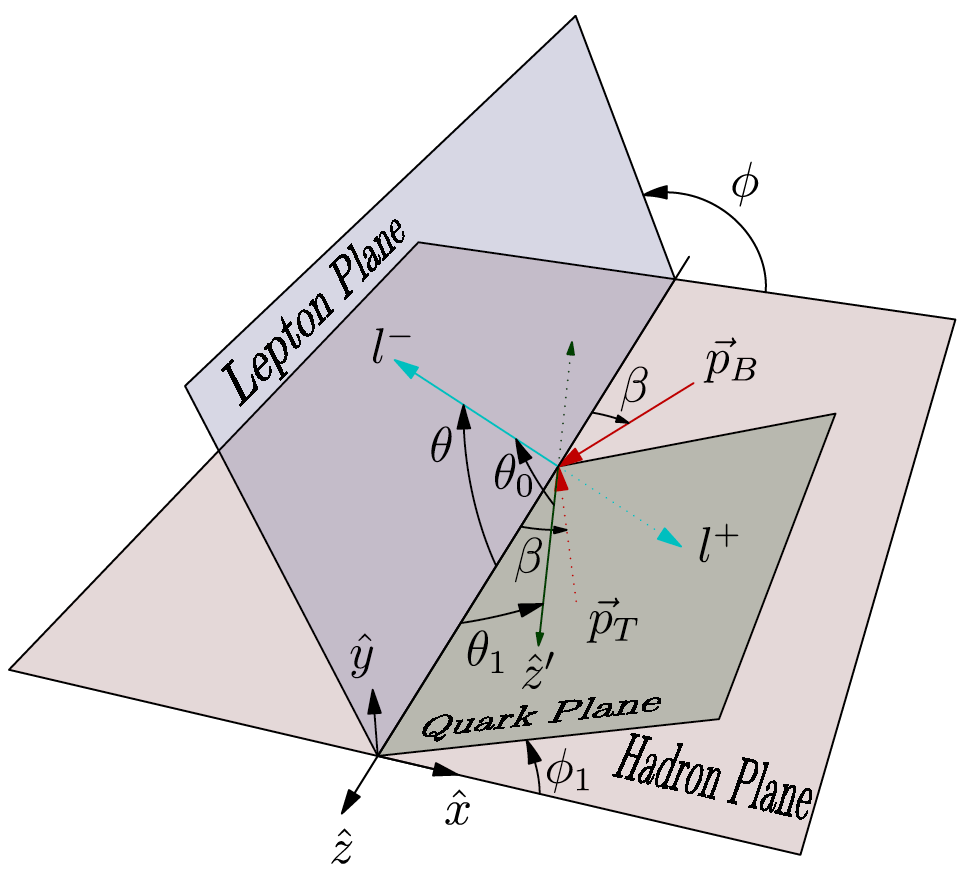}
	\caption{Collins-Soper reference frame}
	\label{fig:geometricalapproach}
\end{figure}
The main idea of this approach is that in the rest frame of a virtual photon $\gamma^{*}$ or $Z$-boson, one can define the dedicated axis directed along the momentum of the annihilating quark (or antiquark). Along this so-called "natural axis", the angular distribution of leptons has axial symmetry and has a simple form
\begin{align}
	   \frac{d\sigma}{d\Omega} \propto 1 + a \cos\theta_0 + \cos^2\theta_0,
	   \label{formula:geom_distr}
\end{align}
where $\theta_0$ is the angle between the $\ell^{-}$ momentum and this natural axis, and the parameter $a$ accounts for the forward-backward asymmetry. 
One can see from the fig.\ref{fig:geometricalapproach} that the quark axis, determined by the direction of its momentum, generally does not coincide with the $z$ axis and forms angle $\theta_1$ with it. In addition, the plane formed by the quark axis and the $z$ axis (the "quark plane") can be rotated by the azimuth angle $\phi_1$. Thus, it is possible to associate the natural axis with the CS reference frame and express all information about the angular distributions in terms of the geometric parameters $\theta_1$ and $\phi_1$. Noting that $\cos{\theta_0}$ satisfies the following relation:
\begin{align}
	\cos\theta_0 = \cos\theta \cos\theta_1 + \sin\theta \sin\theta_1 \cos(\phi - \phi_1),
\end{align}
and substituting it into (\ref{formula:geom_distr}), we obtain a decomposition that exactly matches the general expression for angular distribution:
\begin{align}
	\nonumber
	\frac{d\sigma}{d\Omega} &= \frac{3}{16\pi} [ 1 + \cos^2\theta + \frac{\sin^2\theta_1}{2} (1 - 3\cos^2\theta) + \frac{1}{2} \sin 2\theta_1 \cos\phi_1 \sin 2\theta \cos\phi  \\ \nonumber
	&+ \frac{1}{2} \sin^2\theta_1 \cos 2\phi_1 \sin^2\theta \cos 2\phi 
	+ a \sin\theta_1 \cos\phi_1 \sin\theta \cos\phi + a \cos\theta_1 \cos\theta + \cdots ]
\end{align}
where one can denote 
\begin{gather}
	\nonumber
	A_0 = \langle \sin^2\theta_1 \rangle, \quad
	A_1 = \tfrac{1}{2}\langle \sin 2\theta_1 \cos\phi_1 \rangle, \quad
	A_2 = \langle \sin^2\theta_1 \cos 2\phi_1 \rangle, \\ 
	A_3 = \langle a \sin\theta_1 \cos\phi_1 \rangle, \quad
	A_4 = \langle a \cos\theta_1 \rangle,
	\label{eq:geometrical_coeffs}
\end{gather}
as follows~\cite{Chang:2017kuv,Peng:2019boj}.
Thus, the angular coefficients $A_i$ are related to the orientation of the quark axis with respect to the CS reference frame. 
One can also note that in CS frame the angle $\beta$ between $\hat{z}$ and the hadron momenta is related to the transverse momentum:
\begin{align}
	\tan\beta = \frac{Q_T}{Q}, \qquad
	\sin\beta = \frac{Q_T}{\sqrt{Q^2 + Q_T^2}}, \qquad
	\cos\beta = \frac{Q}{\sqrt{Q^2 + Q_T^2}}.
\end{align}
In the case of quark-antiquark annihilation subprocess angle $\sin\theta_1 = \sin\beta$ and in case of Compton scattering $\sin\theta_1 \approx \sqrt{5}Q_T/\sqrt{Q^2 + 5 Q^2_T}$. As a result, all angular coefficients $A_i$ are given by linear combinations of each subprocess:
\begin{align}
	A_0 &= f \frac{Q_T^2}{(Q^2 + Q_T^2)} + (1 - f)\frac{5Q_T^2}{(Q^2 + 5Q_T^2)}, \quad A_2 = A_0 r_2 \\
	A_1 &= r_1 \left[ f \frac{Q_T Q}{Q^2 + Q_T^2} + (1 - f) \frac{\sqrt{5}\, Q_T Q}{Q^2 + 5Q_T^2} \right], \\
	A_3 &= r_3 \left[ f \frac{Q_T}{(Q^2 + Q_T^2)^{1/2}} + (1 - f) \frac{\sqrt{5}\, Q_T}{(Q^2 + 5Q_T^2)^{1/2}} \right], \\
	A_4 &= r_4 \left[ f \frac{Q}{(Q^2 + Q_T^2)^{1/2}} + (1 - f) \frac{Q}{(Q^2 + 5Q_T^2)^{1/2}} \right].
\end{align}
Here $f$ denotes the relative contribution of the $q\bar{q}$-annihilation subprocess, while factors $r_i$ account phenomenologically for effects that are not contained in the simple coplanar partonic structure. 
	\section{Results of the comparison}
	This section presents a comparison of the theoretical predictions of three approaches for the angular distributions of the Drell-Yan process with experimental data from LHC obtained by ATLAS~\cite{ATLAS:2016rnf}, CMS~\cite{CMS:2015cyj,CMS:2026amb} and LHCb~\cite{LHCb:2022tbc} collaborations.	The comparison is primarily focused on the interval $Q_T < 10$ GeV, where TMD factorization is expected to be applicable. At larger $Q_T$, fixed-order pQCD becomes increasingly appropriate. The geometrical approach provides an alternative phenomenological description based primarily on kinematic considerations. For the collinear approach, the result obtained in~\cite{Lyubovitskij:2024jlb}, which includes both the full pQCD calculation (including NNLP corrections) and the LP approximation is used. For TMD factorization the results of Ref.~\cite{Piloneta:2024aac} were reproduced with the  $\mathtt{artemide}$ program available from~\cite{artemide}. The calculation uses parameters of the unpolarized TMDPDF from global fit ART23~\cite{Moos:2023yfa}, as well as parameters for the Boer-Mulders function given below the eq.~(\ref{eq:boer_mulders}). The uncertainty band for the TMD approach includes variations induced by a selected set of MSHT20~\cite{Bailey:2020ooq} PDF replicas and uncertainties associated with the Boer-Mulders parametrization. The latter contribution is estimated as the standard deviation of the predictions obtained using the central and varied Boer-Mulders parameter sets. For the geometric model, the parameters that were extracted from the LHC experimental data on the angular coefficients and summarized in Table~1 of Ref.~\cite{Lyubovitskij:2025oig} are used. In particular, the relation $A_2 = r_2 \cdot A_0$ is evaluated using $r_2 = 0.661 \pm 0.02$ obtained from the ATLAS $A_2$ data in the central rapidity region $0 < |y| < 1$. The three approaches involve different levels of phenomenological input. While collinear calculation uses globally fitted PDFs, its parameters are not adjusted to the angular coefficients. In contrast, the Boer-mulders parameters entering in TMD calculations and the parameters for geometrical approach were extracted from angular coefficients datasets. Therefore, the values of $\chi^2/N_{\text{pts}}$ presented below should be interpreted primarily as a quantitative measure of the agreement with the data. All calculations for collinear and TMD approaches are performed at $\sqrt{s} = 8$ TeV in the range $80 < Q < 100$  GeV, for three rapidity ranges: central $|y| < 1$ and forward $1 < |y| < 2$, $2 < |y| < 3.5$.
	\begin{table}[h!]
	\centering
	\caption{Comparison of estimated $\chi^2/N_{\text{pts}}$ at $Q_T < 10$ GeV and $0 < |y| < 1$}
	\begin{tabular}{c|c|c|c|c}
		\hline
		\hline
		Angular coefficient & TMD factorization & Full pQCD & LP pQCD & Geometrical approach \\
		\hline
		$A_0$ & $1.131$ & $1.963$ & $1.533$ & $1.288$ \\
		\hline
		$A_1$ & $2.587$ & $2.425$ & $8.409$ & $2.387$ \\
		\hline
		$A_2$ & $2.028$ & $3.862$ & $4.659$ & $3.403$ \\
		\hline
		$A_3$ & $5.661$ & $5.283$ & $29.338$ & $5.388$ \\
		\hline
		$A_4$ & $4.090$ & $3.213$ & $33.821$ & $3.862$ \\
		\hline
		\hline
	\end{tabular}
	\label{tab:chi_sq}
	\end{table}
	TMD factorization systematically provides the better overall description of the low-$Q_T$ data on $A_0$ and $A_2$ angular coefficients, whereas the LP approximation in collinear factorization is insufficient for quantitative description of most angular coefficients.
	\begin{align}
		\chi^2/N_{\text{pts}} = \dfrac{1}{N_{\text{pts}}} \sum\limits^{N_{\text{pts}}}_{i=1} \dfrac{(A^{\text{th}}_i - A^{\text{exp}}_i)^2}{\sigma^{2}_{i,\text{exp}}},
		\label{eq:chi_sq}
	\end{align}
	here $N_{\text{pts}}$ is the number of experimental points with $Q_T < 10$ GeV, and $\sigma_{i,\text{exp}}$ denotes the experimental total uncertainties.
	\subsection{$A_0$,$A_2$ angular coefficients and the Lam-Tung relation}
\begin{figure}[h!]
		\centering
		\begin{minipage}{0.32\textwidth}
			\centering
			\includegraphics[width=\textwidth]{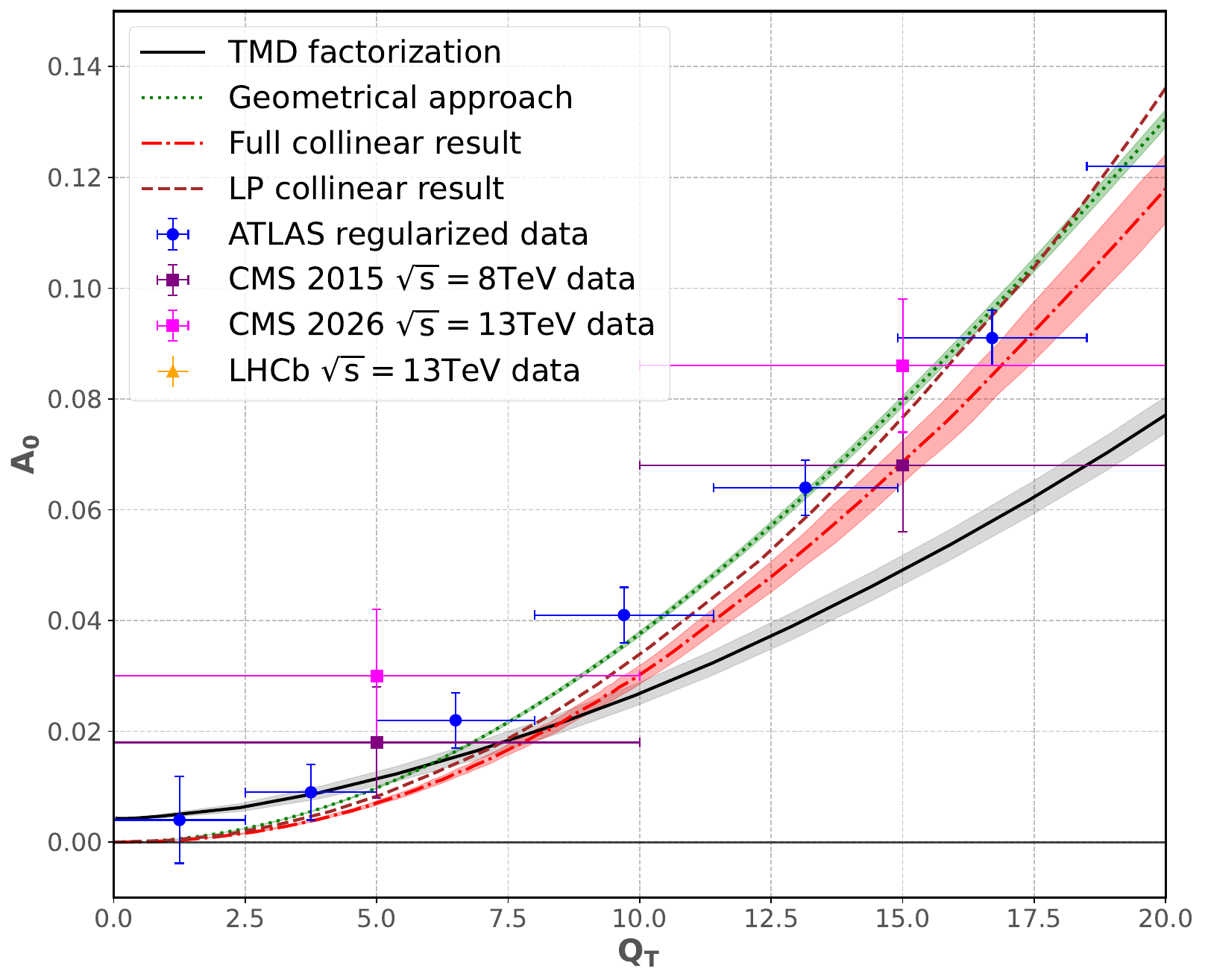}
			\subcaption*{a) $0 < |y| < 1$}
			\label{fig:a0_y01}
		\end{minipage}
		\begin{minipage}{0.32\textwidth}
			\centering
			\includegraphics[width=\textwidth]{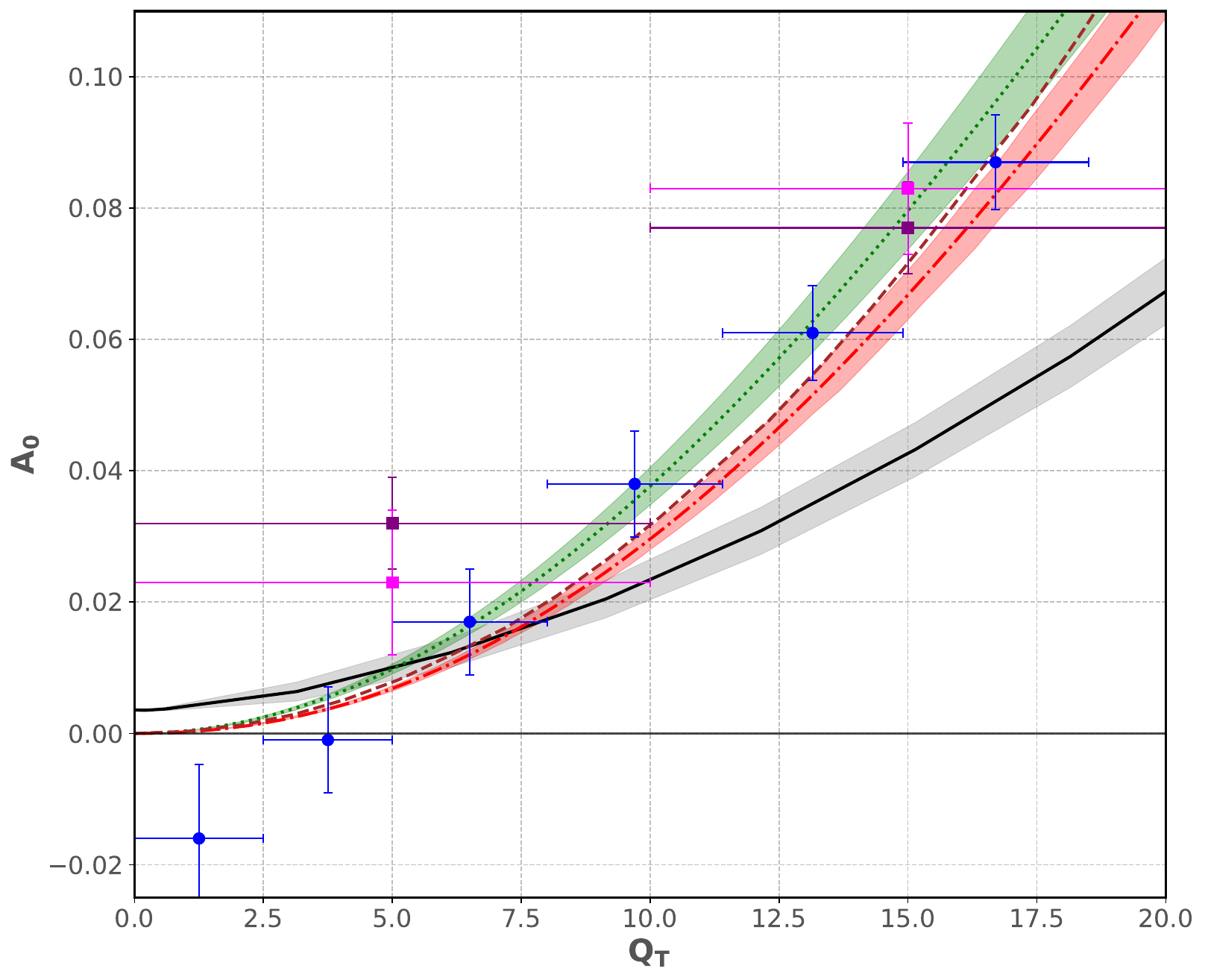}
			\subcaption*{b) $1 < |y| < 2$}
			\label{fig:a0_y12}
		\end{minipage}
		\begin{minipage}{0.32\textwidth}
			\centering
			\includegraphics[width=\textwidth]{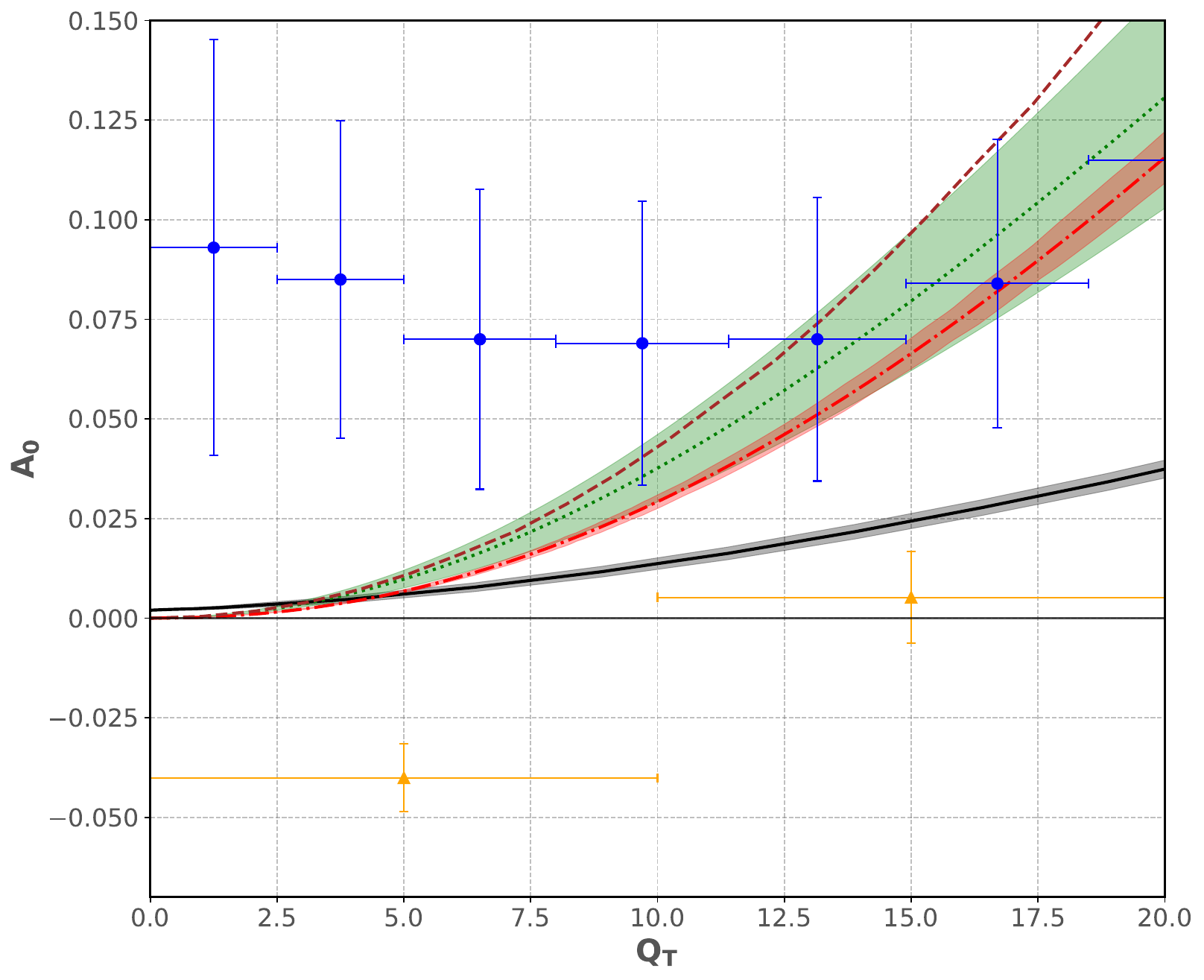}
			\subcaption*{c) $|y| > 2$}
			\label{fig:a0_y235}
		\end{minipage}
		\caption{Results for $A_0$ angular coefficient}
		\label{fig:a0_combined}
	\end{figure}

\begin{figure}[h!]
		\centering
		\begin{minipage}{0.32\textwidth}
			\centering
			\includegraphics[width=\textwidth]{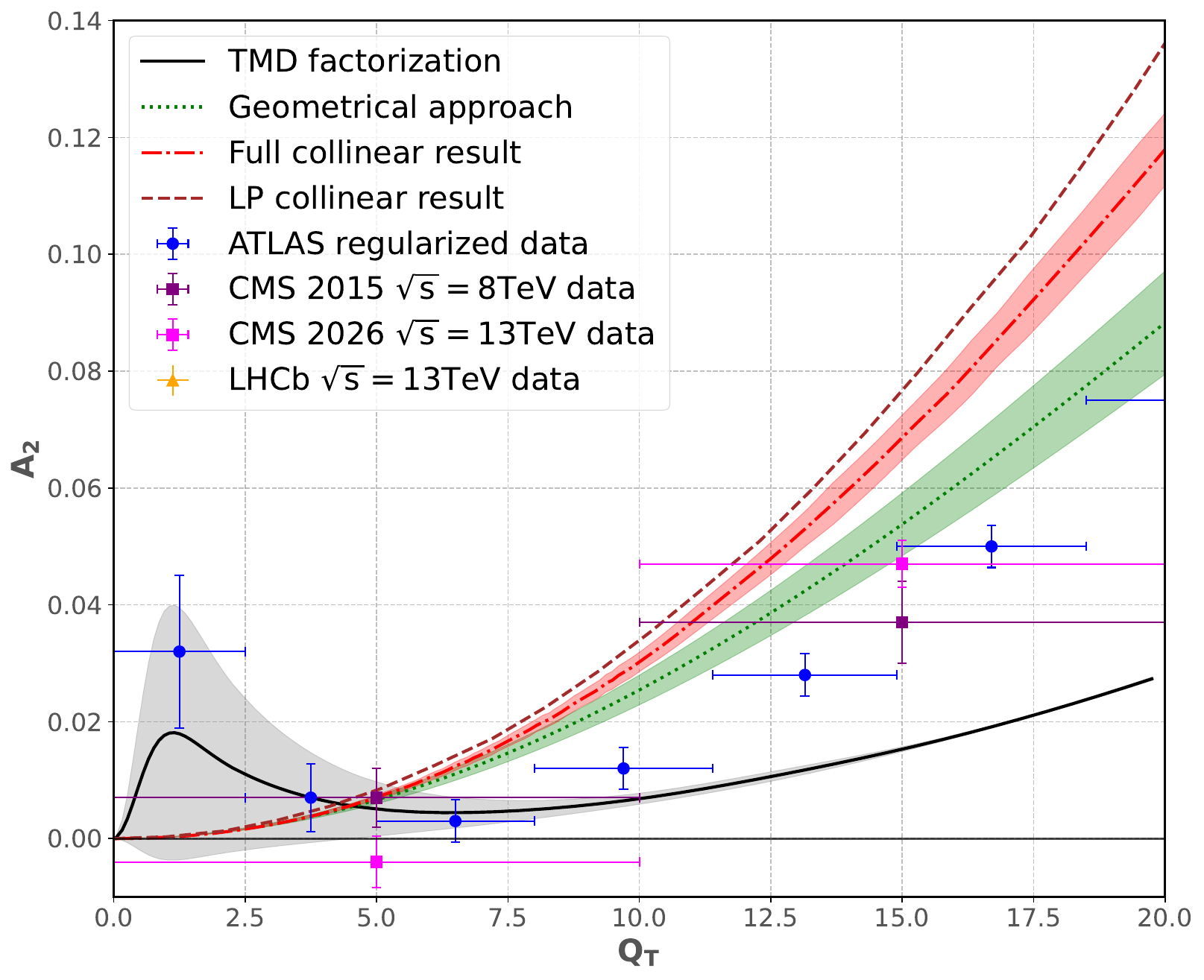}
			\subcaption*{a) $0 < |y| < 1$}
			\label{fig:a2_y01}
		\end{minipage}
		\begin{minipage}{0.32\textwidth}
			\centering
			\includegraphics[width=\textwidth]{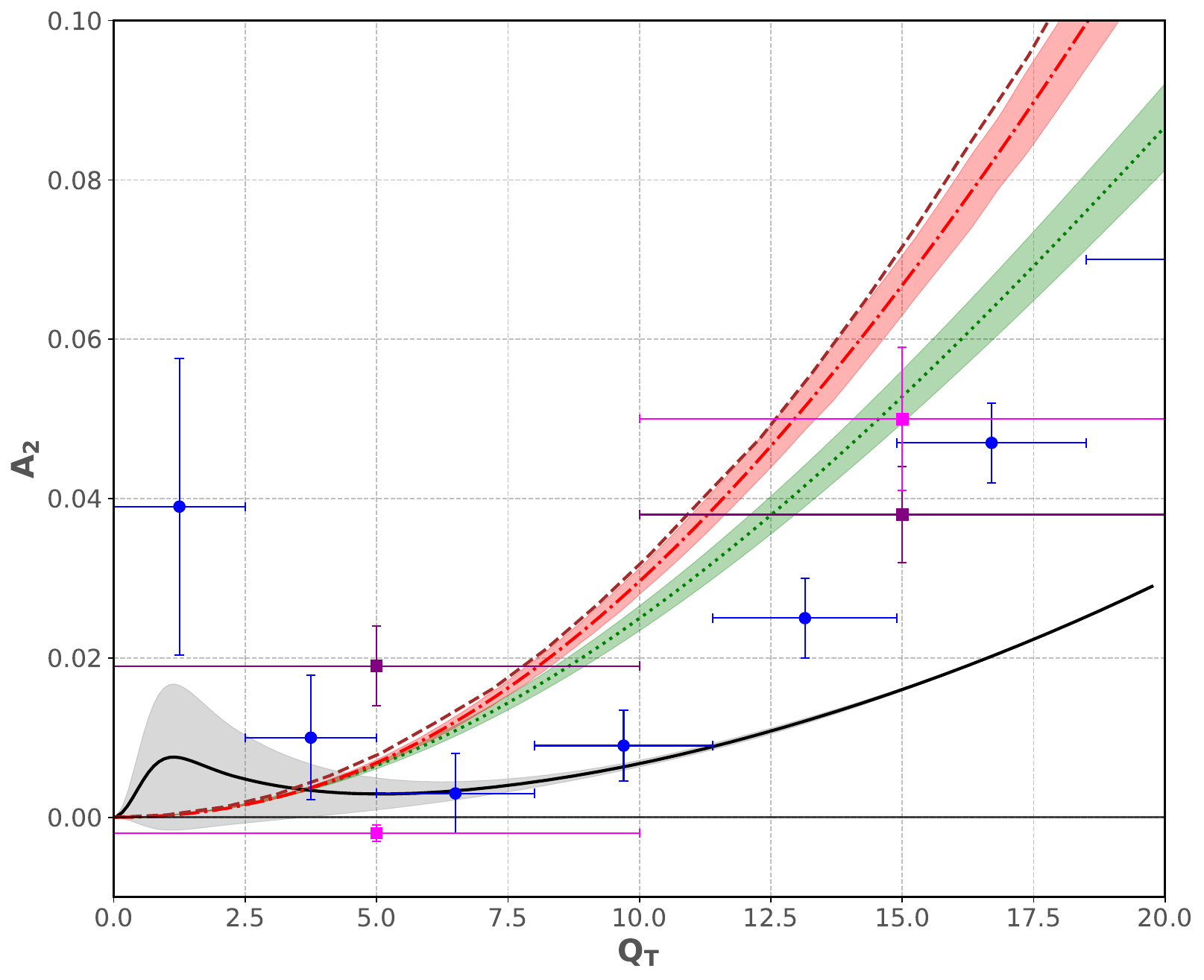}
			\subcaption*{b) $1 < |y| < 2$}
			\label{fig:a2_y12}
		\end{minipage}
		\begin{minipage}{0.32\textwidth}
			\centering
			\includegraphics[width=\textwidth]{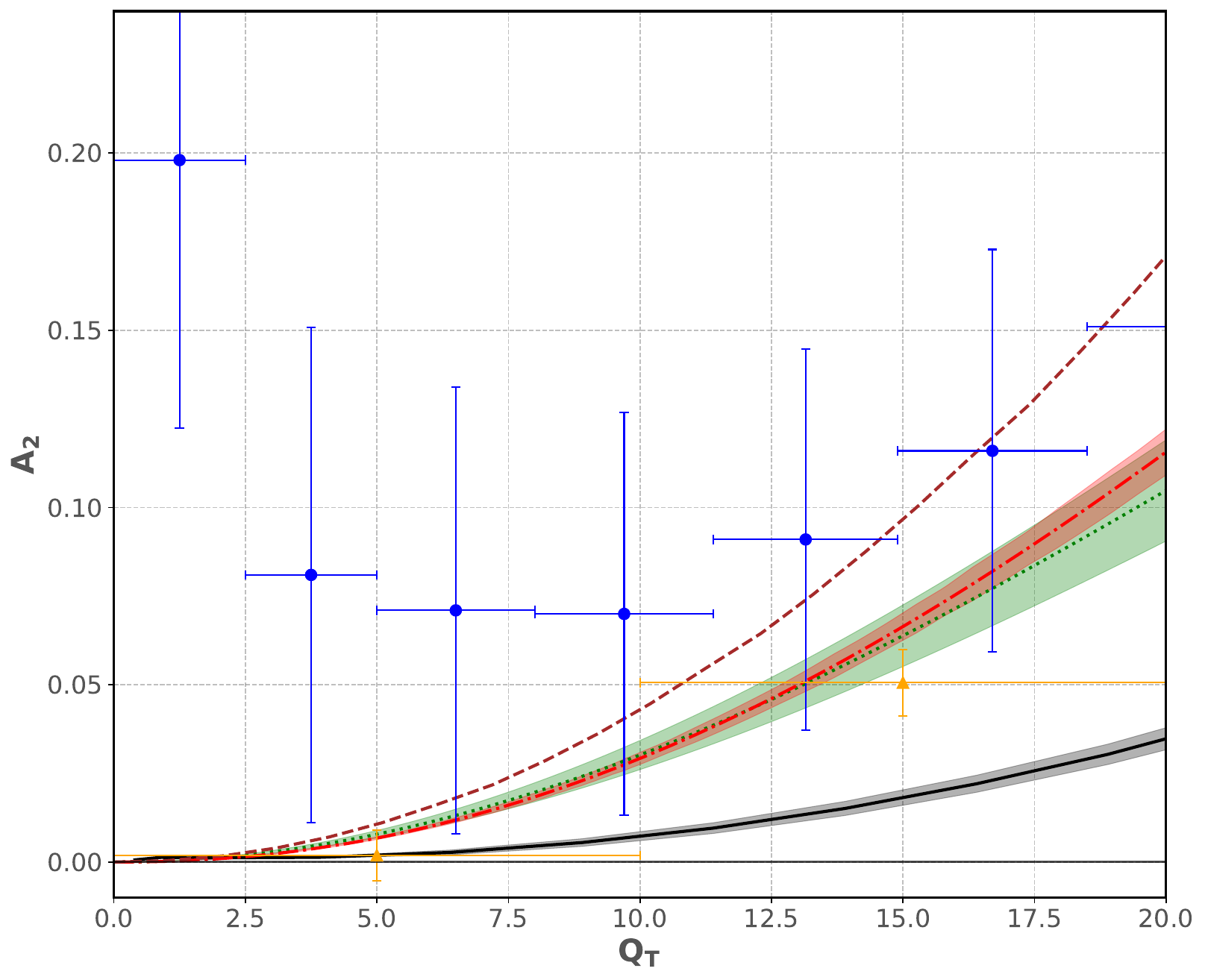}
			\subcaption*{c) $|y| > 2$}
			\label{fig:a2_y235}
		\end{minipage}
		\caption{Results for $A_2$ angular coefficient}
		\label{fig:a2_combined}
	\end{figure}

\begin{figure}[h!]
		\centering
		\begin{minipage}{0.32\textwidth}
			\centering
			\includegraphics[width=\textwidth]{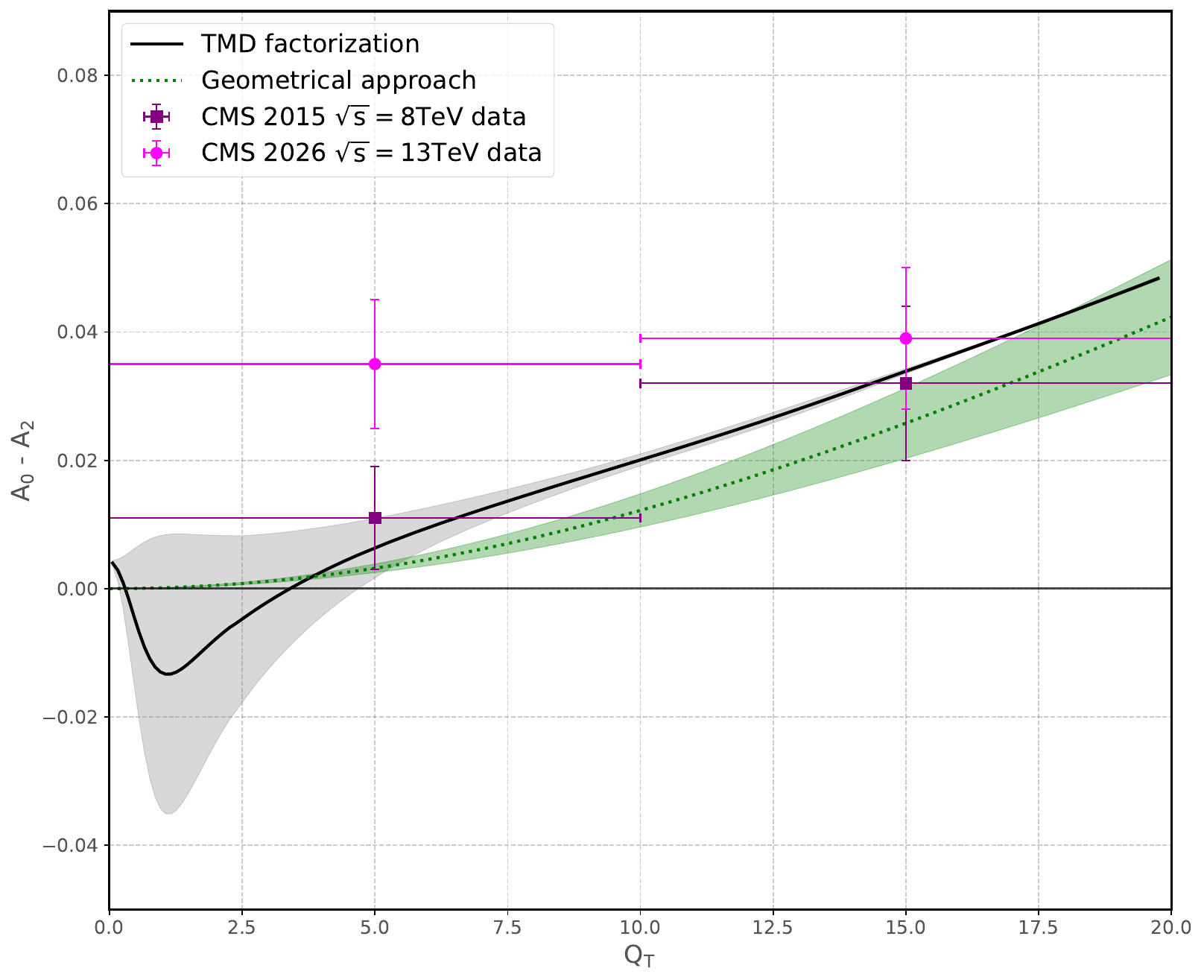}
			\subcaption*{(a)}
			\label{fig:LT_y01}
		\end{minipage}
		\begin{minipage}{0.32\textwidth}
			\centering
			\includegraphics[width=\textwidth]{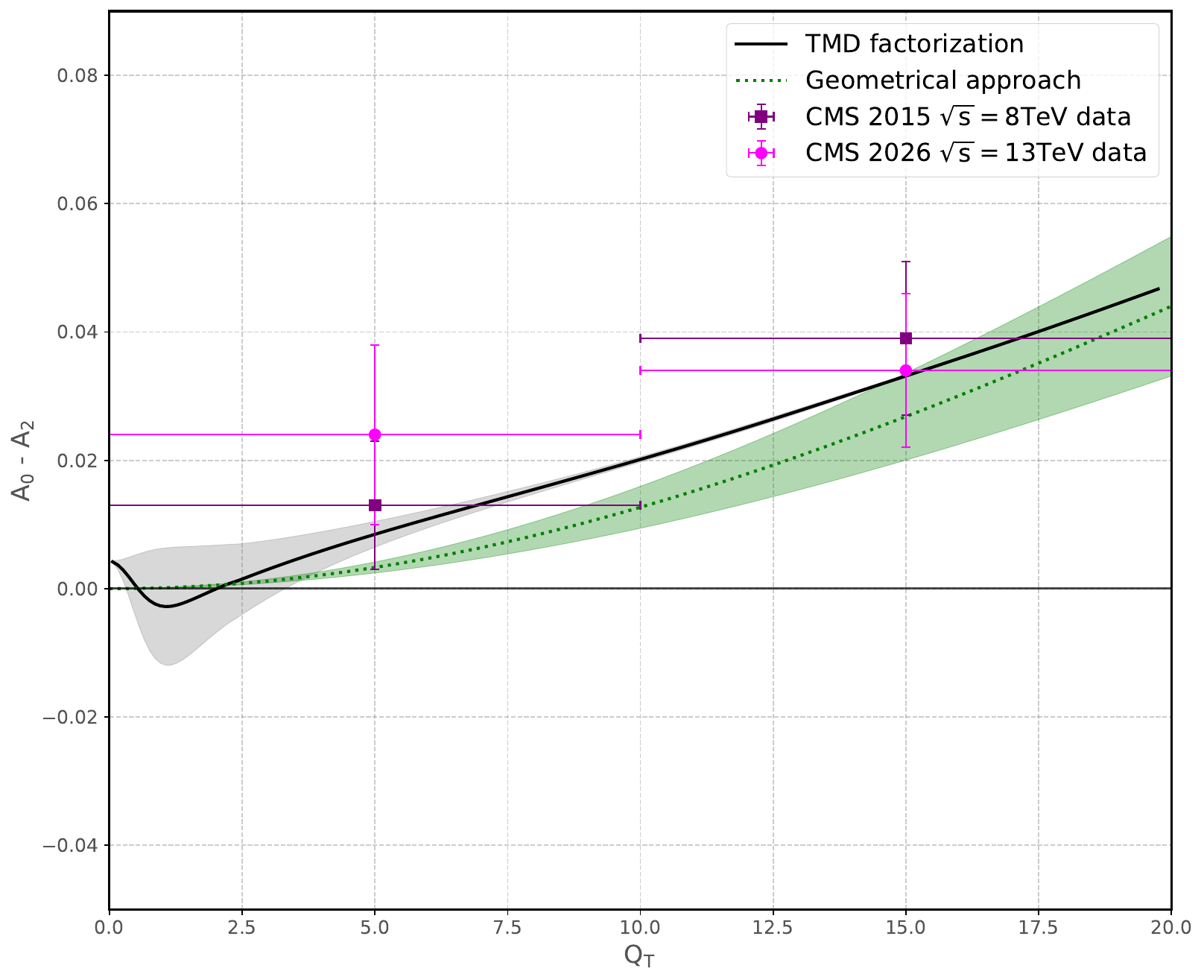}
			\subcaption*{(b)}
			\label{fig:LT_y12}
		\end{minipage}
		\begin{minipage}{0.3\textwidth}
			\centering
			\includegraphics[width=\textwidth]{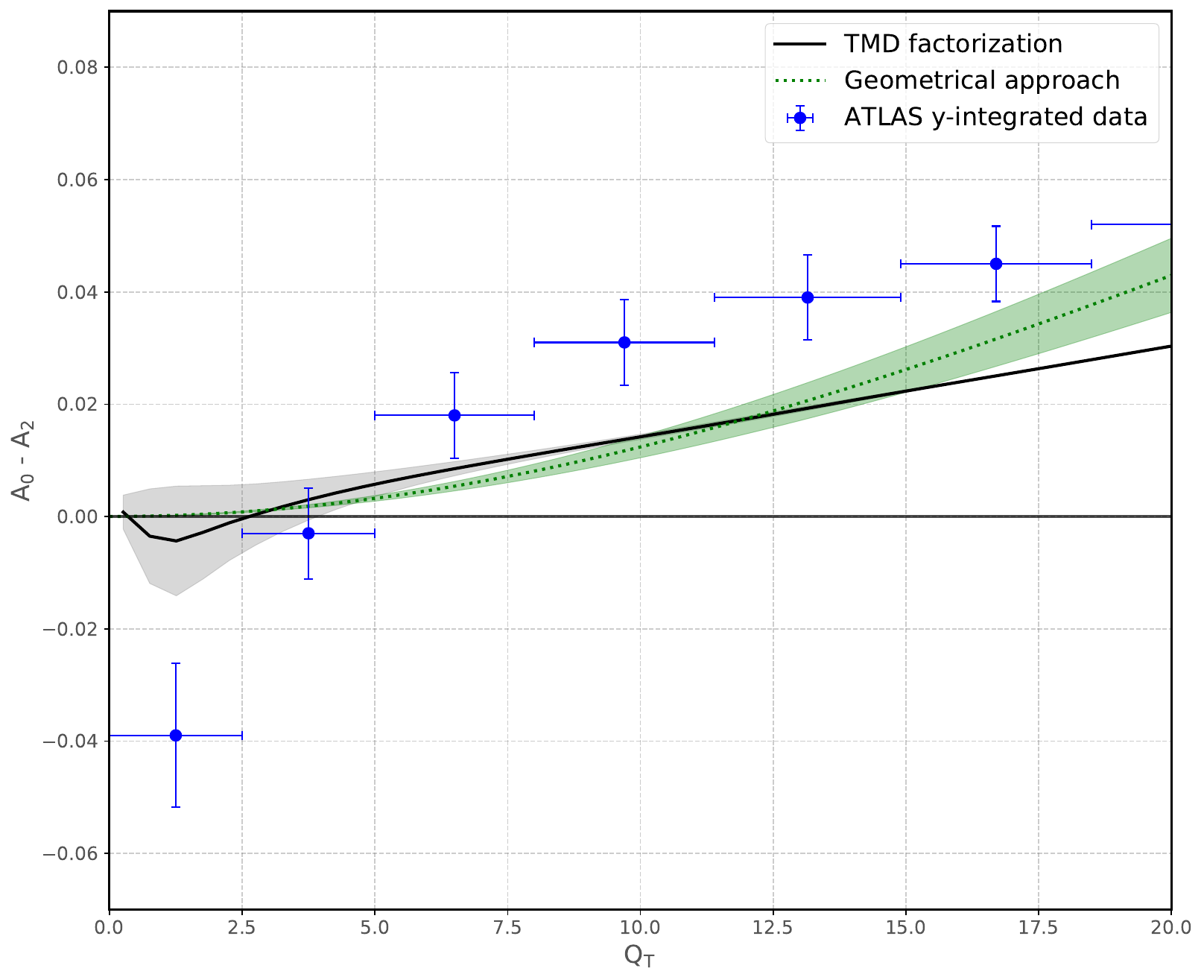}
			\subcaption*{(c)}
			\label{fig:LY_yint}
		\end{minipage}
		\caption{Results for Lam-Tung relation: comparison of theoretical predictions with experimental data in different rapidity regions a) $0 < |y| < 1$, b) $1 < |y| < 2$, c) $|y|-$integrated}
		\label{fig:LT_combined}
	\end{figure}

	Figure~\ref{fig:a0_combined} and Figure~\ref{fig:a2_combined} show the results of comparing the angular coefficient $A_0$ and $A_2$ with the available LHC experimental data. One can see from figures and Table~\ref{tab:chi_sq} that the prediction of TMD approach describes the experimental data better than the other approaches at $Q_T < 10$ GeV in central rapidity region. However, it deviates substantially from the data at $Q_T > 10$ GeV, while the other approaches provide better agreement. At small transverse momentum, fixed-order collinear calculations contain large logarithmic contributions $\ln{(Q^2/Q^2_T)}$, which require all-order resummation. These logarithms are naturally resummed within TMD factorization, explaining its agreement with experimental data in this kinematic region. As the transverse momentum increases, the logarithmic enhancement becomes less important and the fixed-order collinear calculation becomes more appropriate. It is worth noting that in $|y| > 2$ rapidity region the low-$Q_T$ measurements of ATLAS and LHCb for $A_0$ exhibit qualitatively different behaviour. The ATLAS data remain positive in this region, whereas one LHCb point has a negative central value, which lies outside the physical positivity bound $A_0 \geqslant 0$. 
	The key feature of $A_2$ angular coefficient in TMD factorization is that the contribution of the Boer-Mulders function is not suppressed in eq.~(\ref{eq:TMD_S2}) by power of $Q^2$. This leads to the characteristic behavior of $A_2$ at very small $Q_T < 5$ GeV, where the contribution from $h^{\perp}_{1}$ produces a peak. Together with the KPC contribution, this effect leads to a violation of the Lam-Tung relation. In the NLO ($\alpha_s$) collinear factorization the Lam-Tung relation is fulfilled, predicting $A_0=A_2$, the violation of which occurs only at NNLO ($\alpha^2_s$). In geometrical model it follows directly from the acoplanarity between the quark and hadron planes. FIG.~\ref{fig:LT_combined} shows the Lam-Tung relation in TMD and geometrical approaches against experimental data. Both approaches show good agreement of calculations in the $0 < |y| < 1$ and $1 < |y| < 2$ regions with the experimental data. For the $y$-integrated data, however, the TMD prediction underestimates the measurements, with the discrepancy mainly associated with the large-$|y|$ region. This behaviour should not be interpreted as a limitation of the TMD factorization itself. Rather, it reflects the presently available parametrization of the Boer-Mulders function, whose description at large-x has already been discussed in Ref.~\cite{Piloneta:2024aac}. It may also be related to  higher-twist contributions, either kinematical or dynamical. At low-$Q_T$, both the Boer-Mulders contribution and KPCs affect the Lam-Tung relation. This behaviour is consistent with the TMD analysis of Ref.~\cite{Piloneta:2024aac}. Here, the comparison is extended to collinear pQCD and the geometrical approach using the same experimental dataset.
	\subsection{$A_1$, $A_3$, $A_4$ angular coefficients}
\begin{figure}[h!]
		\centering
		\begin{minipage}{0.32\textwidth}
			\centering
			\includegraphics[width=\textwidth]{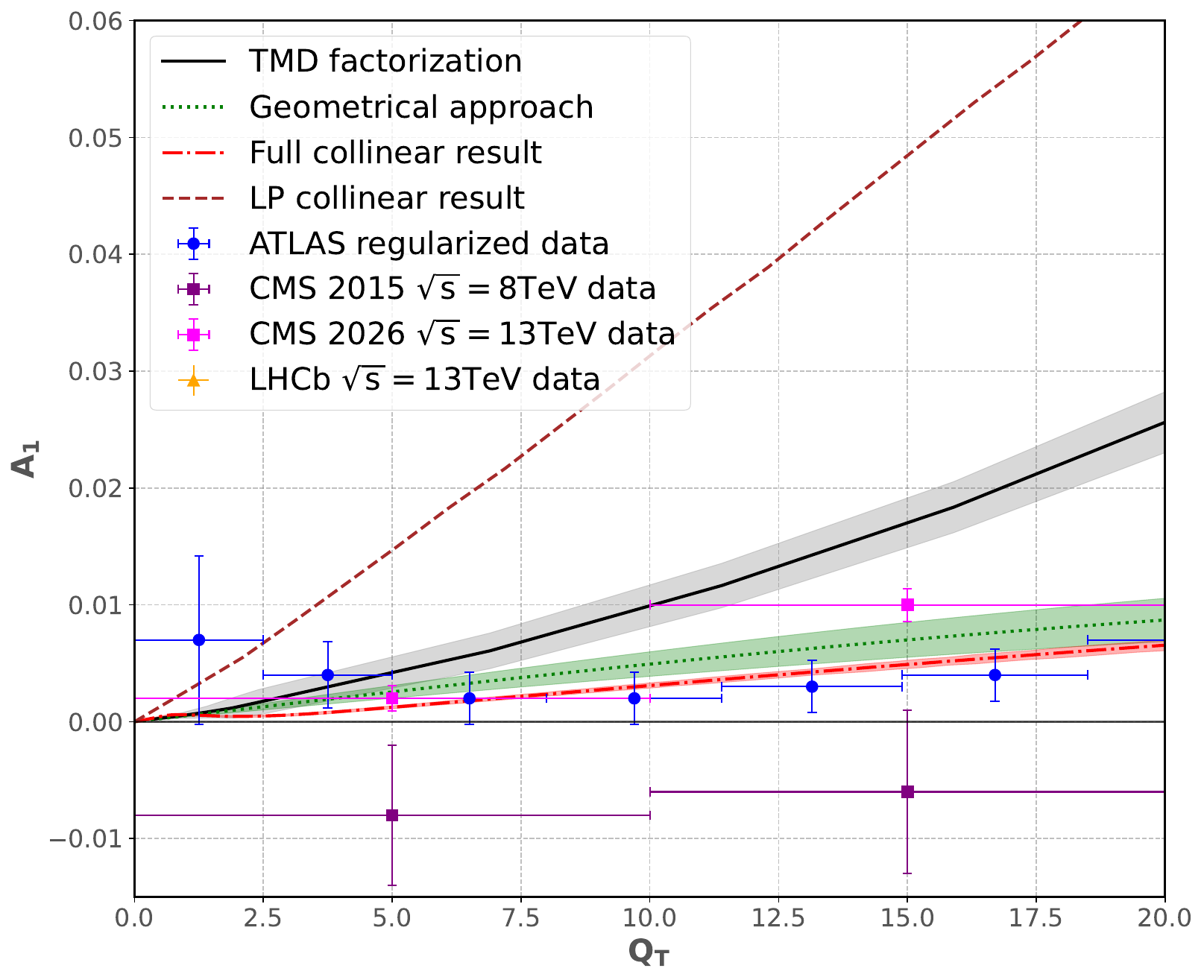}
			\subcaption*{a) $0 < |y| < 1$}
			\label{fig:a1_y01}
		\end{minipage}
		\begin{minipage}{0.32\textwidth}
			\centering
			\includegraphics[width=\textwidth]{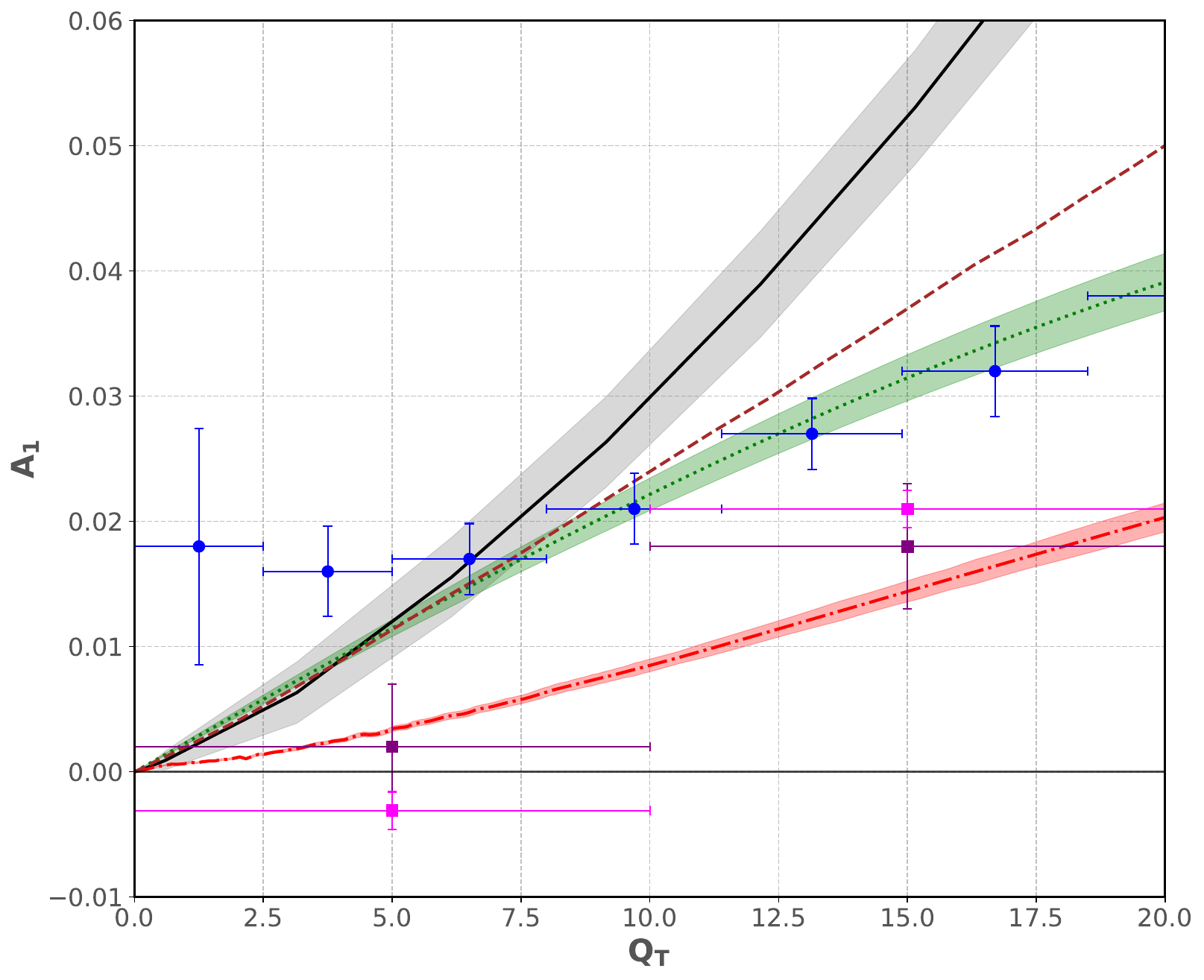}
			\subcaption*{b) $1 < |y| < 2$}
			\label{fig:a1_y12}
		\end{minipage}
		\begin{minipage}{0.32\textwidth}
			\centering
			\includegraphics[width=\textwidth]{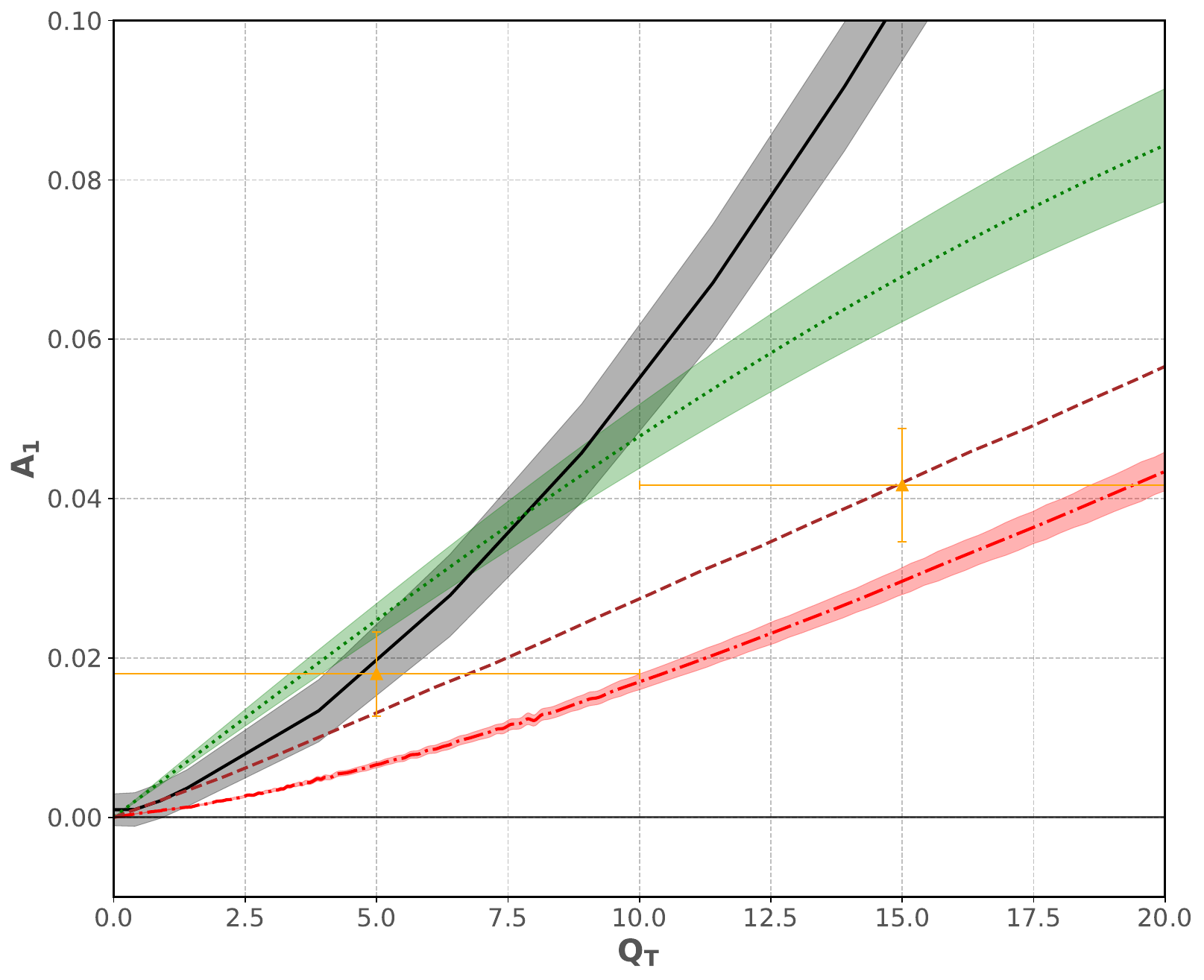}
			\subcaption*{c) $|y| > 2$}
			\label{fig:a1_y235}
		\end{minipage}
		\caption{Results for $A_1$ angular coefficient}
		\label{fig:a1_combined}
	\end{figure}

\begin{figure}[h!]
		\centering
		\begin{minipage}{0.32\textwidth}
			\centering
			\includegraphics[width=\textwidth]{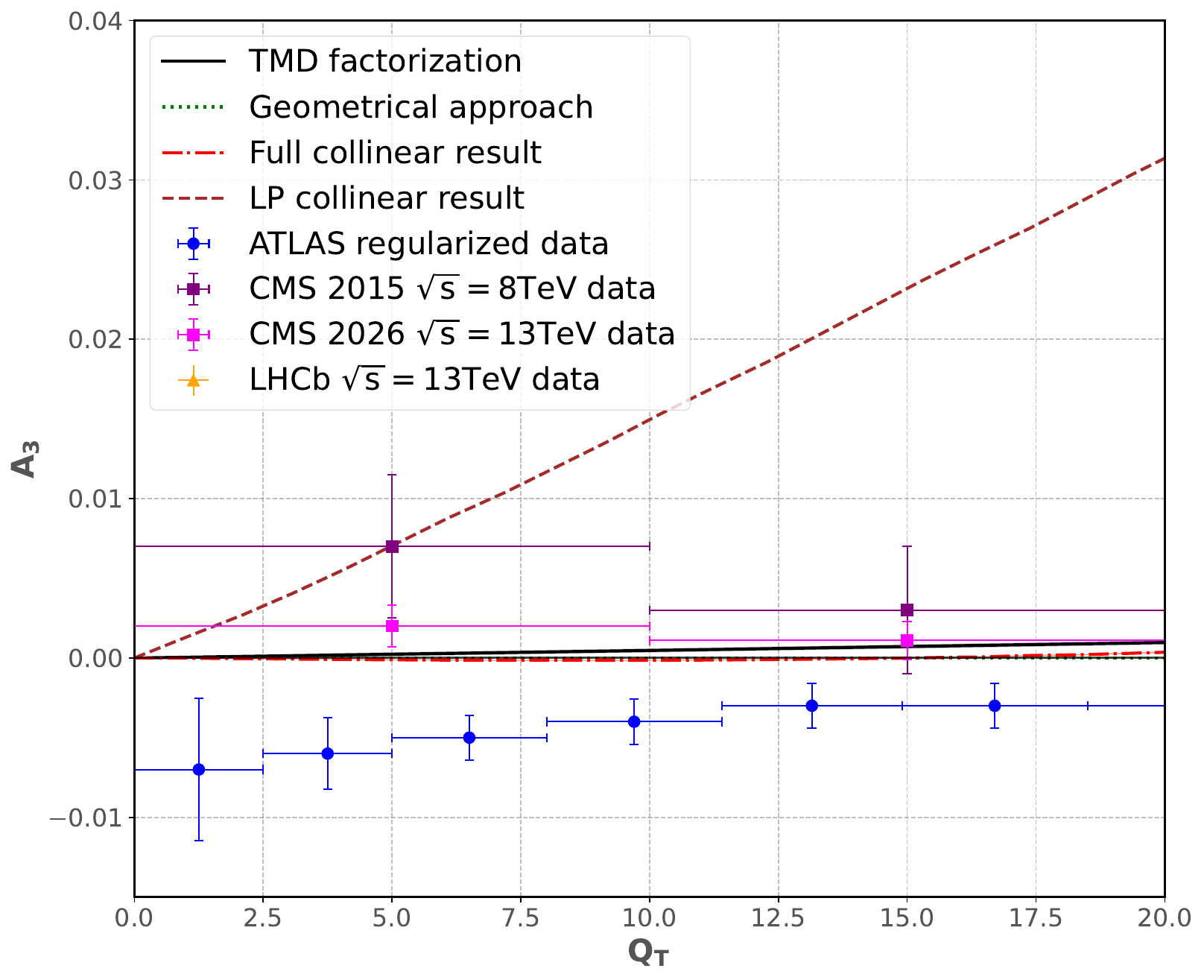}
			\subcaption*{a) $0 < |y| < 1$}
			\label{fig:a3_y01}
		\end{minipage}
		\begin{minipage}{0.32\textwidth}
			\centering
			\includegraphics[width=\textwidth]{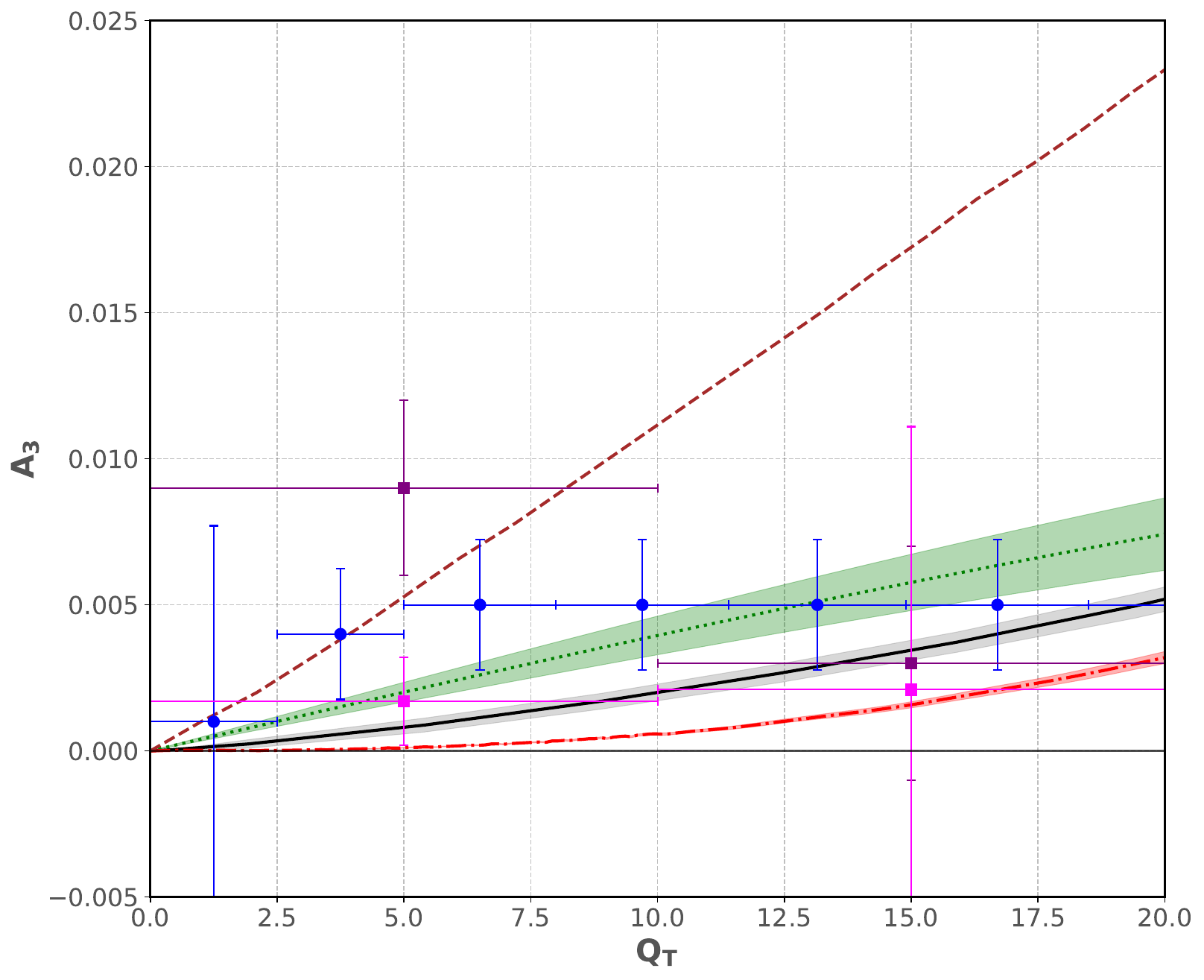}
			\subcaption*{b) $1 < |y| < 2$}
			\label{fig:a3_y12}
		\end{minipage}
		\begin{minipage}{0.32\textwidth}
			\centering
			\includegraphics[width=\textwidth]{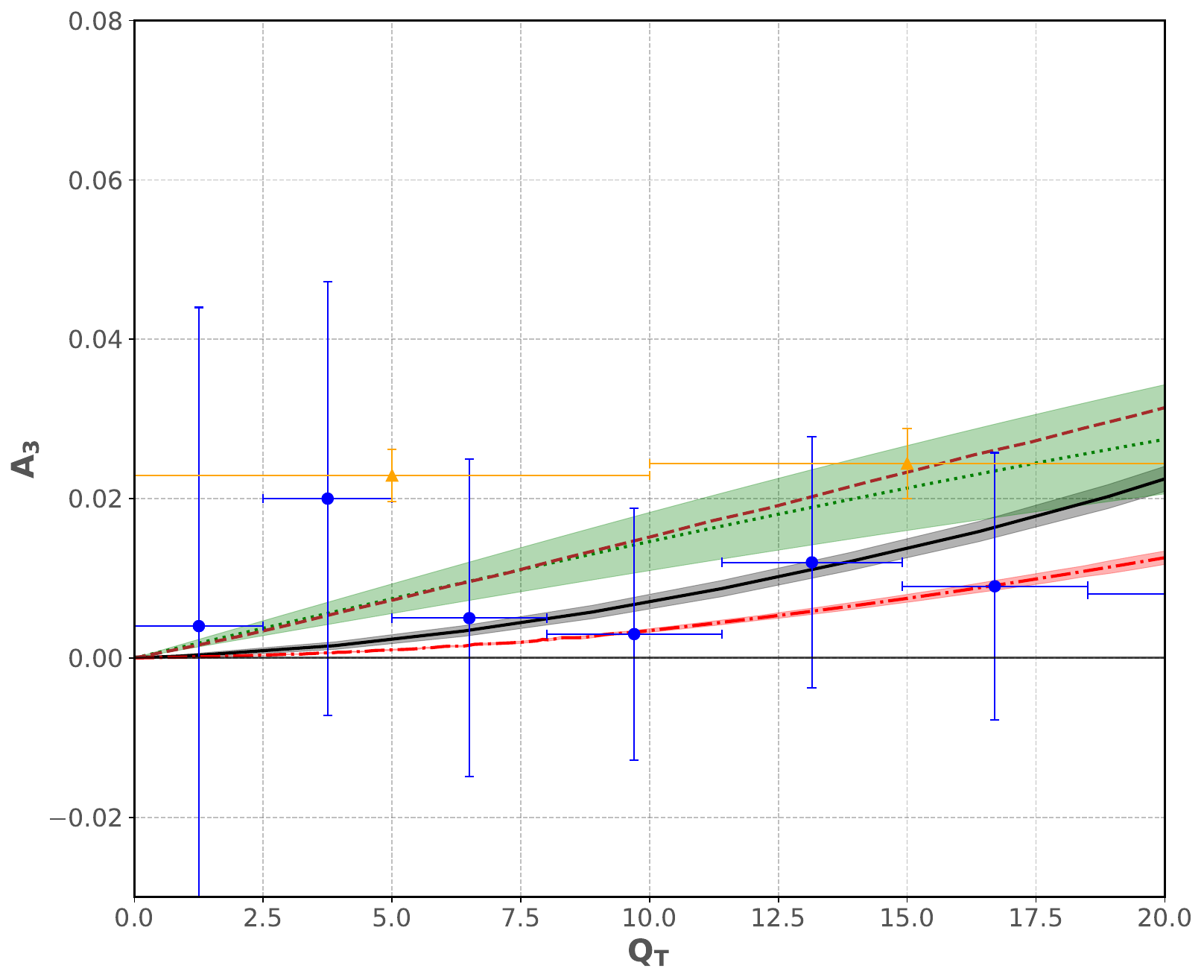}
			\subcaption*{c) $|y| > 2$}
			\label{fig:a3_y235}
		\end{minipage}
		\caption{Results for $A_3$ angular coefficient}
		\label{fig:a3_combined}
	\end{figure}

\begin{figure}[h!]
		\centering
		\begin{minipage}{0.32\textwidth}
			\centering
			\includegraphics[width=\textwidth]{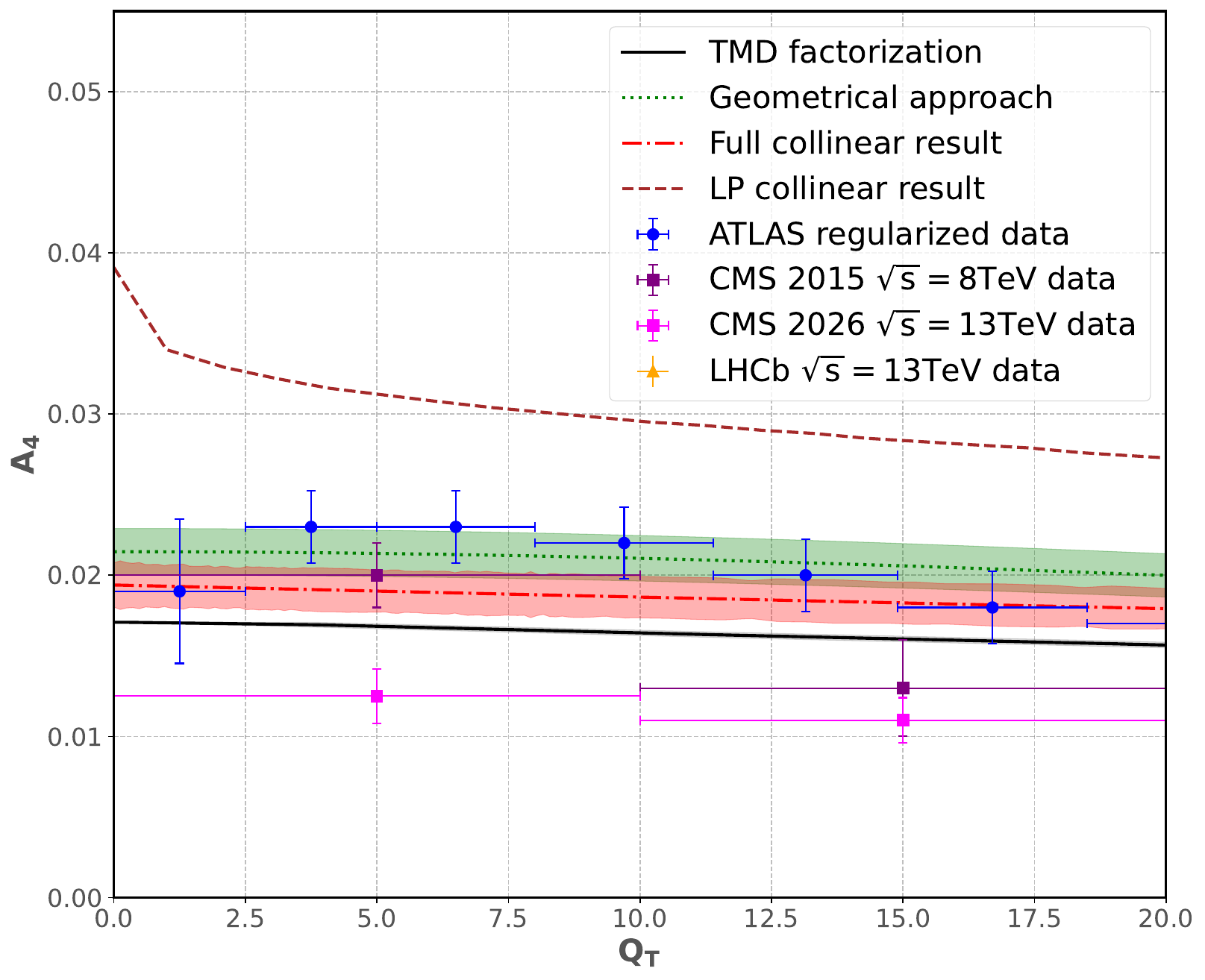}
			\subcaption*{a) $0 < |y| < 1$}
			\label{fig:a4_y01}
		\end{minipage}
		\begin{minipage}{0.32\textwidth}
			\centering
			\includegraphics[width=\textwidth]{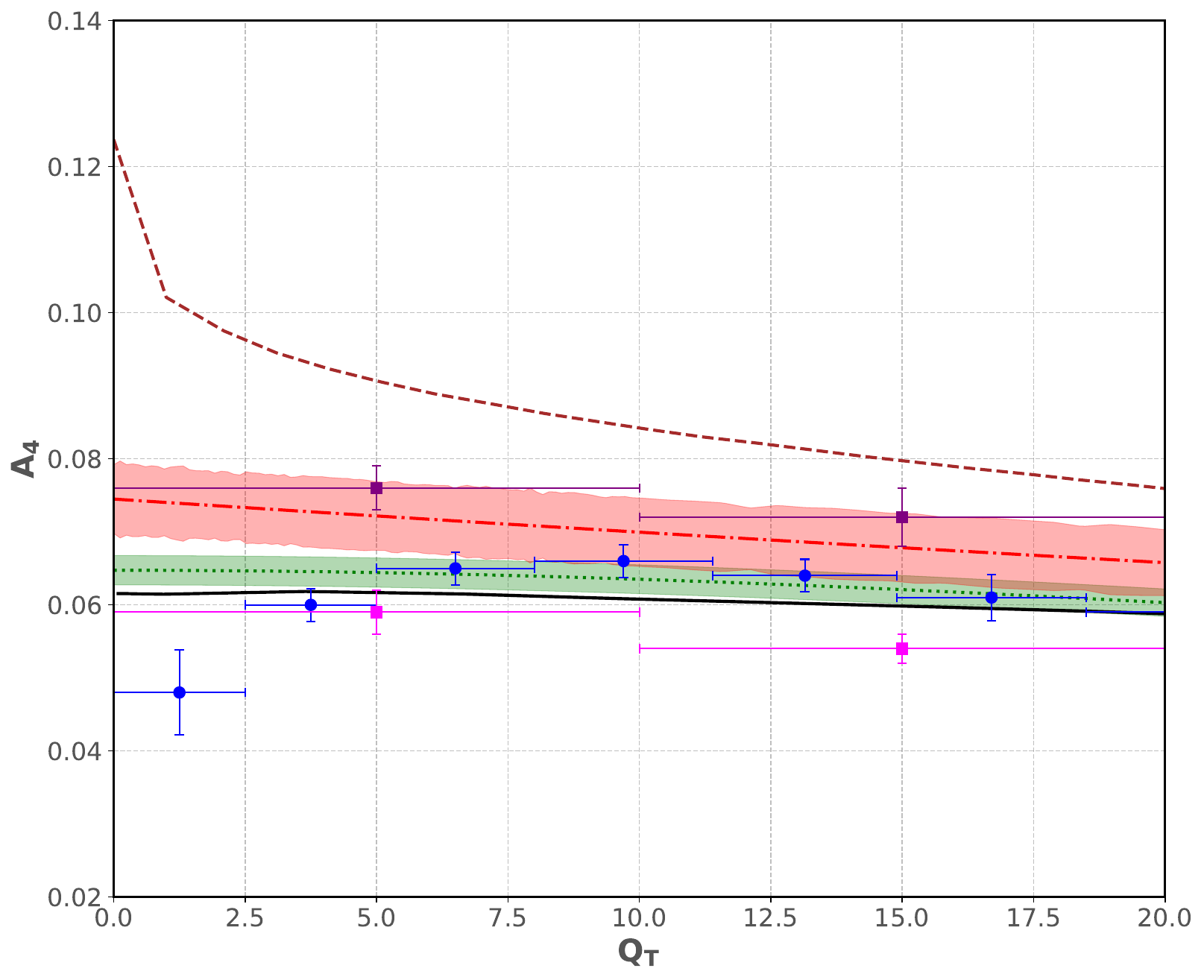}
			\subcaption*{b) $1 < |y| < 2$}
			\label{fig:a4_y12}
		\end{minipage}
		\begin{minipage}{0.32\textwidth}
			\centering
			\includegraphics[width=\textwidth]{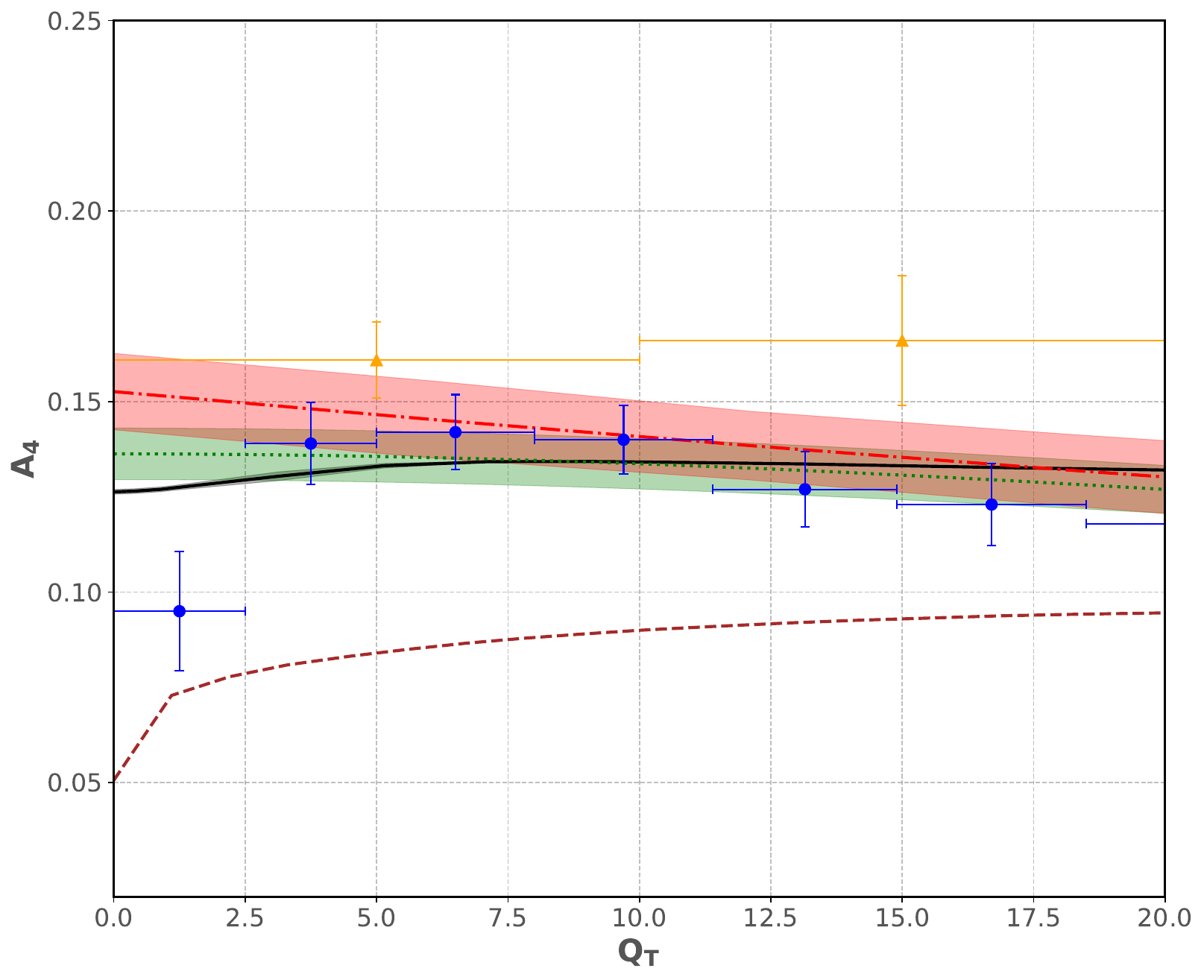}
			\subcaption*{c) $|y| > 2$}
			\label{fig:a4_y235}
		\end{minipage}
		\caption{Results for $A_4$ angular coefficient}
		\label{fig:a4_combined}
	\end{figure}

	Unlike the Lam-Tung relation, these coefficients are expected to be less sensitive to transverse momentum effects. Their comparison provides complementary information on the differences between the three approaches.
	Figures~\ref{fig:a1_combined}, ~\ref{fig:a3_combined} and ~\ref{fig:a4_combined} show the results of comparing the angular coefficient $A_1$, $A_3$ and $A_4$ with the available LHC experimental data. The differences between the theoretical predictions are generally smaller for these coefficients than for $A_0$ and $A_2$, although the quantitative agreement with the data varies between coefficients and experimental datasets. In central rapidity region the ATLAS data differ significantly from the CMS data and include negative values of the $A_3$ coefficient, resulting in an increased value of the $\chi^2/N_{\text{pts}}$ shown in Table~\ref{tab:chi_sq}. The results of the calculations for the $A_1$ angular coefficient show a reasonably good agreement with the data in the low-$Q_T$ region. The result for the $A_4$ angular coefficient gives a similar picture, but without strong deviation between TMD factorization and other approaches in the region $Q_T > 10$ GeV. The relatively small differences between the predictions for the $A_3$ and $A_4$ coefficients indicate that these observables are considerably less sensitive to the details of the factorization formalism than the Lam-Tung combination. In particular, the stability of the $A_4$ coefficient reflects the robustness of the forward-backward asymmetry, which is largely determined by electroweak couplings rather than by QCD dynamics. 
	\section{Conclusion}
	We have compared three theoretical approaches to the description of angular distributions in DY $Z$-boson production. The predictions of TMD factorization, collinear pQCD and the geometrical model were compared with the latest high-precision measurements for energies $\sqrt{s} = 8$ and $13$ TeV from the ATLAS, CMS and LHCb collaborations. 
	The comparison shows that the differences between the three approaches are most pronounced for the $A_0$ and $A_2$ coefficients, which are particularly sensitive to transverse-momentum effects. In the central rapidity region at $Q_T < 10$ GeV, the TMD factorization provides the most accurate description of these coefficients. The Lam-Tung relation is also well described by TMD and geometrical approaches in the central and intermediate rapidity regions, where transverse momentum effects and partonic acoplanarity are relevant. At larger rapidities, $|y| > 2$, the description deteriorates for both $A_0$ and $A_2$, which may indicate additional higher-twist contributions, either kinematical or dynamical.
	For $A_1, A_3$ and $A_4$ all three approaches are generally closer to one another, and no single approach provides a uniformly superior description. The $A_4$ coefficient is comparatively stable with respect to the choice of framework, consistent with its dominant sensitivity to electroweak couplings. The LP calculations for collinear factorization significantly deviate from the experimental data and predictions of other approaches, whereas the inclusion of higher-power terms greatly improves the description. Overall, the comparison shows that the relative performance of the three approaches depends strongly on the kinematic region and on the angular coefficients considered. Among the observables considered, $A_0, A_2$ and the Lam-Tung relation show the strongest sensitivity to transverse momentum dynamics.  
	\section*{Acknowledgments}
	The authors are grateful to J.-C. Peng, A. Vladimirov, and A. Zhevlakov for valuable discussions and comments.

\end{document}